\documentclass[journal,11pt,onecolumn,draftclsnofoot,a4paper]{IEEEtran}

\usepackage{graphicx}

\usepackage[noadjust]{cite}
\usepackage{dsfont}
\usepackage{psfrag}
\usepackage[cmex10]{amsmath}
\usepackage{amssymb}
\usepackage{amsthm}
\usepackage{url}
\usepackage{booktabs}
\usepackage{multirow}
\usepackage[caption=false,font=footnotesize]{subfig}

\usepackage{xcolor}
\definecolor{landanimal}{rgb}{.545,.1,.1}
\colorlet{ocean}{blue!60!black}

\newcommand{\EE}{{\mathbb E}}

\newcommand{\T}{\scriptscriptstyle T}

\newcommand{\E}{{\rm E}}
\newcommand{\uM}{{\mathbf u}}

\newcommand{\w}{{\mathbf w}}
\newcommand{\wo}{{\mathbf{w}}_{\rm o}}

\newcommand{\q}{{\mathbf q}}

\newcommand{\wtil}{\widetilde{\mathbf w}}
\newcommand{\wtilt}{\widetilde{{\boldsymbol w}}}
\newcommand{\psitil}{\widetilde{\boldsymbol \psi}}

\newcommand{\Tr}{{\rm Tr}}

\newcommand{\phiM}{{\boldsymbol{\phi}}}
\newcommand{\psiM}{{\boldsymbol{\psi}}}

\newcommand{\col}{{\rm col}}
\newcommand{\diag}{{\rm diag}}
\newcommand{\Mcal}{\boldsymbol{\cal M}}
\newcommand{\Ccal}{\boldsymbol{\cal C}}
\newcommand{\Rcal}{\boldsymbol{\cal R}}
\newcommand{\Bcal}{\boldsymbol{\cal B}}
\newcommand{\Scal}{\boldsymbol{\cal S}}
\newcommand{\Qcal}{\boldsymbol{\cal Q}}
\newcommand{\Zcal}{\boldsymbol{\cal Z}}
\newcommand{\Lcal}{\boldsymbol{\cal L}}
\newcommand{\Xcal}{\boldsymbol{\cal X}}
\newcommand{\Sgm}{\boldsymbol{\Sigma}}

\begin{document}

\title{Adaptive Diffusion Schemes for Heterogeneous Networks}

\author{Jesus~Fernandez-Bes,~
        Jer\'onimo~Arenas-Garc\'{i}a,~
        Magno~T.~M.~Silva,~
        and~Luis~A.~Azpicueta-Ruiz,~
\thanks{The work of Fernandez-Bes was supported by projects TIN2013-41998-R and DPI2016-75458-R from Spanish Ministry of Economy and Competitiveness (MINECO), Spain, MULTITOOLS2HEART from CIBER-BBN through Instituto de Salud Carlos III, Spain, European Social Fund (EU) and Arag\'on Government through BSICoS group (T96) and by the European Research Council (ERC) through project ERC-2014-StG 638284. The work of Arenas-Garc\'{\i}a was partly supported by MINECO project TEC2014-52289-R and Comunidad de Madrid project PRICAM S2013/ICE-2933. The work of Silva was partly supported by CNPq under Grant 304275/2014-0 and FAPESP under Grant~\mbox{2012/24835-1}. The work of Luis A. Azpicueta-Ruiz is partially supported by Comunidad de Madrid under grant ’CASI-CAM-CM’ (id. S2013/ICE-2845), by the Spanish Ministry of Economy and Competitiveness (under grant DAMA (TIN2015-70308-REDT) and grant TEC2014-52289-R), and by the European Union.}
\thanks{Fernandez-Bes is with BSICoS Group, I3A, IIS Arag\'on, University of Zaragoza and CIBER-BBN, Zaragoza, 50018, Spain. Arenas-Garc\'{\i}a, and Azpicueta-Ruiz are with Dept.~Signal Theory and Communications, Universidad Carlos III de Madrid, Legan\'es, 28911, Spain. Silva is with Dept.~Electronic Systems Engineering, Escola Polit\'ecnica, Universidade de S\~ao Paulo, S\~ao Paulo, 05508-010, Brazil.}
\thanks{Corresponding author: Jesus~Fernandez-Bes: jfbes@unizar.es}}


\maketitle

\begin{abstract}
In this paper, we deal with distributed estimation problems in diffusion networks with heterogeneous nodes, i.e., nodes that either implement different adaptive rules or differ in some other aspect such as the filter structure or length, or step size. Although such heterogeneous networks have been considered from the first works on diffusion networks, obtaining practical and robust schemes to adaptively adjust the combiners in different scenarios is still an open problem. In this paper, we study a diffusion strategy specially designed and suited to heterogeneous networks. Our approach is based on two key ingredients: 1) the adaptation and combination phases are completely decoupled, so that network nodes keep purely local estimations at all times; and 2) combiners are adapted to minimize estimates of the network mean-square-error. Our scheme is compared with the standard Adapt-then-Combine scheme and theoretically analyzed using energy conservation arguments. Several experiments involving networks with heterogeneous nodes show that the proposed decoupled Adapt-then-Combine approach with adaptive combiners outperforms other state-of-the-art techniques, becoming a competitive approach in these scenarios.
\end{abstract}

\begin{IEEEkeywords}
Adaptive networks, diffusion networks, distributed estimation,
least-squares, mean-square performance.
\end{IEEEkeywords}

%
\IEEEpeerreviewmaketitle

\section{Introduction}
\label{sec:intro}
Over the last years, adaptive diffusion networks
have become an attractive and robust approach to estimate a set of parameters of
interest in a distributed manner (see, e.g., \cite{Lopes2008,Schizas2009,chouvardas2012sparsity,sayed2012diffusion,Sayed_SPM2013, xia2011adaptive,sayin2014compressive, Khalili2012,fadlallah2013diffusion, plata2015distributed, arablouei2014distributed,di2013sparse,di2014diffusion, xu2015distributed,chen2015learning} and their references). Compared to other distributed schemes, such as incremental and consensus strategies, diffusion techniques present some advantages, e.g., they are more robust to link failures or they do not require the definition of a cyclic path that runs across the nodes as in incremental solutions \cite{Lopes2007a}.
Furthermore, they perform better than consensus techniques in terms of stability, convergence rate, and tracking ability \cite{Sayed_SPM2013}. For these reasons, adaptive diffusion networks are considered an efficient solution in applications
such as target localization and tracking \cite{sayed2012diffusion}, environment monitoring \cite{Sayed_SPM2013}, and spectrum sensing in mobile networks \cite{sayed2012diffusion, DiLorenzo2013}, among others. Moreover, they are also suited to model complex behaviors exhibited by biological or socioeconomic networks \cite{Sayed_SPM2013}.

\par Diffusion networks consist of a collection of connected nodes, linked according to a certain topology, that cooperate with each other through local interactions to solve a distributed inference or optimization problem in real time. Each node is able to extract information from its local measurements and combine it with the ones received from its neighbors \cite{sayed2012diffusion,Sayed_SPM2013}.
This is typically performed in two stages: adaptation and combination.
The order in which these stages are performed leads to two possible schemes: Adapt-then-Combine (ATC) and Combine-then-Adapt (CTA) \cite{Lopes2008,Cattivelli2010}. In both cases, the adaptation and combination steps are interleaved with the communication of the intermediate estimates among neighbors. {In general, it is assumed that this communication among the nodes is synchronous, though an analysis of asynchronous diffusion strategies is available in \cite{zhao2015asynchronous}.}
\par {In this paper, we focus on {\it heterogeneous} diffusion networks.\footnote{{Although there exist many studies involving heterogeneous networks in the literature (sensor networks \cite{mhatre2004homogeneous}, epidemiology \cite{moreno2002epidemic} or cellular networks \cite{damnjanovic2011survey}), here we restrict ourselves to the particular case of diffusion networks.}} We refer as {\it heterogeneous} to networks whose nodes implement diverse update functions, i.e., they can differ in the filter length or structure, step sizes, or even in the implemented learning rule. This is different to other popular scenarios such as multitask or node-specific diffusion networks \cite{plata2017heterogeneous}, where the heterogeneity is in the input that nodes receive or/and in the task they solve. Heterogeneous nodes are an interesting choice to improve the tracking performance of the network, or simply to achieve a better tradeoff between computational cost and convergence rate by updating the nodes with different algorithms. Thus, it is not surprising that such heterogeneous networks have been considered in the literature from the first works on diffusion networks. For instance, \cite{sayed2012diffusion} and \cite{Sayed_SPM2013} already considered in the analysis of the ATC and CTA schemes the case of least-mean-squares (LMS) nodes with different step sizes, and {\cite{Tu_TSP2013}} used the term {\it heterogeneous} to refer to informed and uninformed (i.e., without access to local measurements) nodes in a diffusion network.} 

\par
\begin{figure}[t]
	\begin{center}
	\begin{tabular}{c}
	\includegraphics[width=0.8\linewidth]{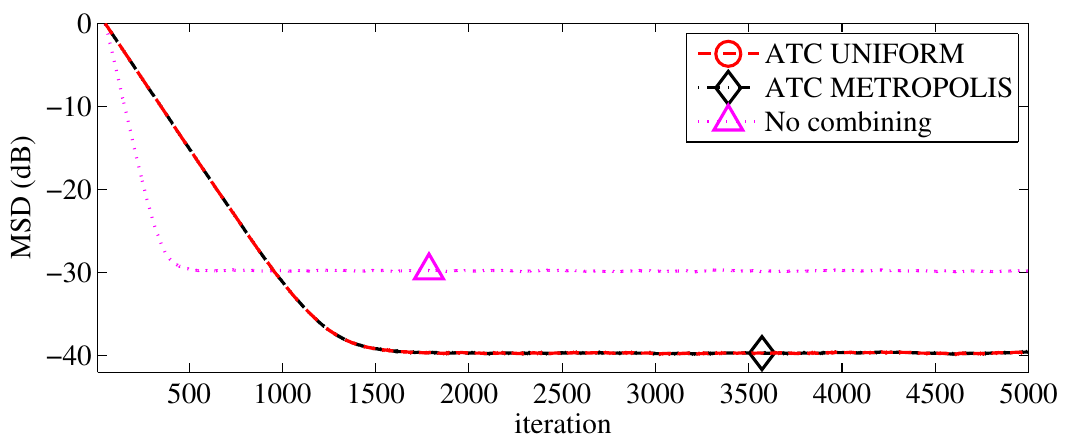} \\ \small{Fast node} \\
	\includegraphics[width=0.8\linewidth]{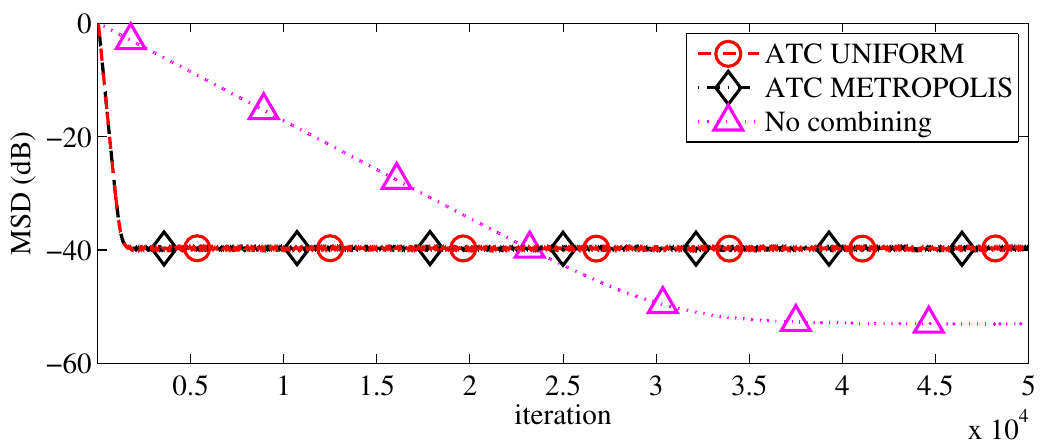} \\ \small{Slow node}
	\end{tabular}
	\end{center}
	\caption{Estimation error for two nodes of a fully-connected ATC network with 5 nodes. All nodes implement LMS rules, but node 1 (fast node) uses a larger step size than the rest of nodes.}
	\label{fig:ATCdiffusion_static}
\end{figure}

 {Compared to multitask scenarios, experimental studies of this kind of networks have been quite limited up to now and we think that they deserve more attention.} One possible explanation is that overall network performance can be very sensitive to an inappropriate selection of the network combiners. To illustrate this, Fig. \ref{fig:ATCdiffusion_static} shows the performance of the ATC scheme of {\cite{Cattivelli2010}} for a fully-connected network of 5 nodes implementing LMS updates, using static combiners. In this example, four nodes have a common step size, whereas the last node uses a larger value that can provide faster convergence. As it can be seen, in this case the diffusion strategy actually degrades the convergence of the fast node and the steady-state performance of the slow ones with respect to the operation of these nodes in the non-cooperation case. This is due to the use of fixed combiners that `contaminate' the estimation of the fast node during convergence, whereas during the steady-state regime slow nodes fuse their estimations with that coming from the fast node, resulting in larger rather than smaller estimation error.

\par {This simple example illustrates the importance of adequate rules for setting and adjusting the combination parameters in heterogeneous networks.} Indeed, combination weights play an essential role in the overall performance of the network. For instance, diffusion least-mean-squares (LMS) strategies can perform similarly to classical centralized solutions when the weights used to combine the neighbors estimates are optimally adjusted \cite{sayed2012diffusion,Takahashi2010,zhao2012performance}. Initially, different static combination rules were proposed such as Uniform \cite{blondel2005convergence}, Laplacian \cite{xiao2004fast}, Metropolis \cite{xiao2004fast}, and Relative Degree \cite{cattivelli2008diffusion}. Some adaptive schemes for adjusting the combination weights (e.g., \cite{Takahashi2010,zhao2012performance,tu2011optimal,yu2013strategy}) have also been proposed in order to optimize the network performance under spatially varying signal-to-noise ratio (SNR). Although these adaptive rules can reduce the steady-state error with respect to static combiners, some experiments show a deterioration in the convergence behavior \cite{yu2013strategy}. Consequently, some schemes propose the use of different rules for transitory and steady-state regimes and include mechanisms to switch from one rule to the other in an online manner \cite{yu2013strategy,fernandez2014adjustment}. {Since all these works are generally optimized assuming homogeneous nodes, an important challenge is related with the fact that all the above-mentioned combination rules result in degraded performance when used with heterogeneous networks, and there are presently no alternatives for dealing with such problem in a general case.}

\par In this work, we focus on an alternative diffusion scheme {especifically designed for heterogeneous networks}. In our approach, firstly proposed in \cite{fernandez2015distributed,fernandez2012novel}, and which will be called Decoupled ATC (D-ATC), the adaptation phase is kept decoupled from the combination phase, i.e., the local estimation of each node is combined with the estimates received from its neighbors, as in standard ATC, but the resulting combined estimation is not fed back into the next adaptation step. This scheme presents a more clear separation between the adaptation and combination phases. {As it will be shown later, this allows us to implement mean-square-error (MSE) based rules for the combination phase which offer an adequate behavior for heterogeneous networks.} With these rules we obtain a significant improvement in convergence and steady-state performance with respect to previous approaches, both in tracking and stationary scenarios. In addition, our proposal seems to be a more natural scheme for asynchronous networks, which are receiving increasing attention \cite{zhao2015asynchronous}.

\par This paper extends our previous works \cite{fernandez2015distributed,fernandez2012novel} in different ways:
\begin{enumerate}

\item We analyze the mean behavior of our diffusion strategy and derive sufficient conditions for the network combiners that guarantee the mean stability of the algorithm.

\item Using energy conservation arguments \cite{sayed2008adaptive}, we derive closed-form expressions for the steady-state mean-square deviation (MSD) of the network and of its individual nodes in a non stationary environment. 

\item {We propose two new rules for adjusting the combination weights: One following a Least-Squares (LS) approach, in the same vein as the one introduced in \cite{fernandez2015distributed,fernandez2012novel}, and one based on the Affine Projection Algorithm (APA).} 

\item Finally, we include detailed simulation work, both for stationary and tracking scenarios, to illustrate the performance of the proposed schemes and to corroborate the theoretical results.
\end{enumerate}

\par {The paper is organized as follows. The general formulation of ATC diffusion strategies for heterogeneous networks, together with the introduction of adaptive combiners, is presented in Section~\ref{sec:heterogeneous}. In Section~\ref{sec:datc}, the Decoupled ATC strategy is proposed, and we theoretically analyze it in Section~\ref{sec:performanceanalysis}. APA and LS-based rules for adapting network combiners are  derived  in Section~\ref{sec:Adaptivecombiners}. Experimental results are provided in Section~\ref{sec:simulations}, and  we close the paper with our main conclusions and some possibilities for future works in Section~\ref{sec:conclusion}.
}
\subsection{Notation}

We use boldface lowercase letters to denote vectors and boldface uppercase letters  to denote matrices.
The superscript $T$ represents the transpose of a matrix or a vector. Depending on the context, ${\mathbf{0}}_N$ represents an $N\times N$ matrix or a length-$N$ column vector with all elements equal to zero, and $\mathds{1}_N$ is an all-ones column vector with length $N$. In addition, to simplify the arguments, we assume that all the variables are real. Table \ref{Tabla_var} summarizes the notation that is used throughout the paper.
\begin{table}[t]
	\centering
	\caption{\label{Tabla_var}Summary of the notation used in the paper}
		\begin{tabular}{ll}
			\toprule
			$N$ ~~~& Number of nodes in the network\\
			${\mathcal{N}}_k$ ~~~& Neighborhood of node $k$, including itself\\
            $N_k$ ~~~& Cardinality of ${\mathcal{N}}_k$\\
			${\bar{\cal N}}_k$ & Neighborhood of node $k$, excluding itself\\
             $\bar{N}_k$ ~~~& Cardinality of ${\bar{\cal N}}_k$\\
      		$\bar{{\bf{b}}}_k$ & Vector with the indexes of all nodes in ${\bar{\cal N}}_k$ \\
			$\bar{{{b}}}_k^{(m)}$ & Index of the $m^\text{th}$ node connected to node $k$ \\
			$\wo(n)$ & Unknown time-varying parameter vector \\
			${\bf{\psiM}}_k(n)$ & Local estimate of $\wo(n)$ (based only\\
            $ $ & on local data at node $k$) \\
			${\bf{w}}_{k}(n)$ & Combined estimate of $\wo(n)$ at node $k$\\
			$\{d_k(n),\uM_k(n)\}$~~& Local desired value and regression vector\\
            $ $ & at node $k$ \\
			$v_k(n)$ & Local noise at node $k$\\
			$y_k(n)$ & Local output of node $k$ \\
			$e_k(n)$ & Local error of node $k$ \\
			$c_{\ell k}(n)$ & Combination weight assigned by node $k$ to\\
            $ $ & the estimate received from node $\ell \in {\cal N}_k$\\
			${\bf{c}}_{k}(n)$ & Vector with combination weights\\
            $ $ & associated to node $k$\\
			$\bar{{\bf{c}}}_{k}(n)$ & Vector with the same entries of ${\bf{c}}_{k}(n)$,\\
            $ $ & excluding $c_{kk}(n)$\\
			\bottomrule
		\end{tabular}
\end{table}

\section{Heterogeneous diffusion networks with MSE-based adaptive combiners}
\label{sec:heterogeneous}
\subsection{ATC and CTA diffusion strategies}

\begin{figure}[t]
\centering
\includegraphics[width=0.8\linewidth]{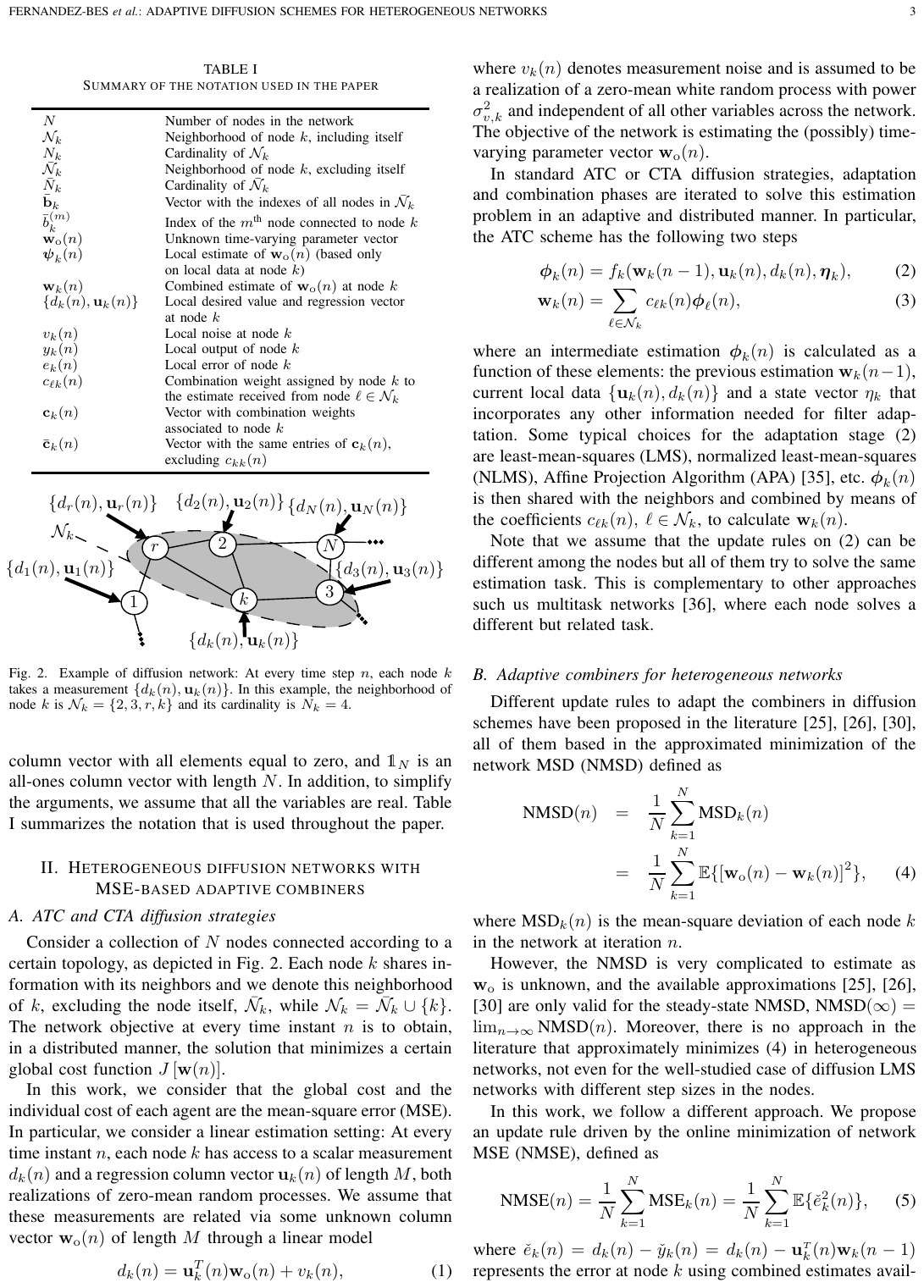}
\caption{Example of diffusion network: At every time step $n$, each node $k$ takes a measurement $\{d_{k}(n),\uM_{k}(n)\}$. In this example, the neighborhood of node $k$ is ${\mathcal{N}}_k=\{2, 3, r, k\}$ and its cardinality is  $N_k=4$.\label{fig:distributed}}
\end{figure}

Consider a collection of $N$ nodes connected according to a certain topology, as depicted in Fig.~\ref{fig:distributed}. Each node $k$ shares information with its neighbors and we denote this neighborhood of $k$, excluding the node itself, $\mathcal{\bar{N}}_k$, while $\mathcal{N}_k = \mathcal{\bar{N}}_k \cup \{k\}$. The network objective at every time instant $n$ is to obtain, in a distributed manner, the solution that minimizes a certain global cost function $J\left[  \w(n)\right]$.

In this work, we consider that the global cost and the individual cost of each agent are the mean-square error (MSE). In particular, we consider a linear estimation setting: At every time instant $n$, each node $k$ has access to a scalar measurement $d_k(n)$ and a regression column vector ${\bf{u}}_{k}(n)$ of length $M$, both realizations of zero-mean random processes. We assume that these measurements are related via some unknown column vector $\wo(n)$ of length $M$ through a linear model
\begin{equation}
d_k(n)=\uM_k^{T}(n)\wo(n)+v_k(n),
\label{eq:dk_model}
\end{equation}
where $v_k(n)$ denotes measurement noise and is assumed to be a realization of a zero-mean white random process with power $\sigma_{v,k}^2$ and independent of all other variables across the network. The objective of the network is estimating the (possibly) time-varying parameter vector $\wo(n)$.


{In standard ATC or CTA diffusion strategies, adaptation and combination phases are iterated to solve this estimation problem in an adaptive and distributed manner. In particular, the ATC scheme has the following two steps
\begin{align}
\phiM_k(n)&= f_k ( \w_k (n-1), \uM_k(n) , d_k(n), \boldsymbol{\eta}_k), \label{eq:diffusion_ATC1}\\
\w_k(n)&= \sum_{\ell\in{\cal N}_k}c_{\ell k}(n)\phiM_{\ell}(n),\label{eq:diffusion_ATC2}
\end{align}
where an intermediate estimation $\phiM_k(n)$ is calculated as a function of these elements: the previous estimation $\w_k(n-1)$, current local data $\{\uM_k(n) , d_k(n)\}$ and a state vector $\mathbf{\eta}_k$ that incorporates any other information needed for filter adaptation. Some typical choices for the adaptation stage \eqref{eq:diffusion_ATC1} are least-mean-squares (LMS), normalized least-mean-squares (NLMS), Affine Projection Algorithm (APA) \cite{sayed2008adaptive}, etc. $\phiM_k(n)$ is then shared with the neighbors and combined by means of the coefficients $c_{\ell k}(n),\:\ell\in{\cal N}_k$, to calculate $\w_k(n)$.}

{Note that we assume that the update rules on \eqref{eq:diffusion_ATC1} can be different among the nodes but all of them try to solve the same estimation task. This is complementary to other approaches such us multitask networks \cite{chen2015diffusion}, where each node solves a different but related task.}

\subsection{Adaptive combiners for heterogeneous networks}
\label{sec:adaptive_combiners_hetero}
%

{Different update rules to adapt the combiners in diffusion schemes have been proposed in the literature \cite{Takahashi2010,tu2011optimal,zhao2012performance}, all of them based in the approximated minimization of the network MSD (NMSD) defined as}
\begin{equation}
\text{NMSD}(n) =  \frac{1}{N} \sum_{k=1}^N   \text{MSD}_k(n) = \frac{1}{N} \sum_{k=1}^N  \EE \{ \left[\wo(n) - \w_k(n)\right]^2\},
\label{eq:NMSDss}
\end{equation}
where $\text{MSD}_k(n)$ is the mean-square deviation of each node $k$ in the network at iteration $n$.

{However, the NMSD is very complicated to estimate as $\wo$ is unknown, and the available approximations \cite{Takahashi2010,tu2011optimal,zhao2012performance} are only valid for the steady-state NMSD, $\text{NMSD}({\infty}) = \lim_{n \rightarrow \infty}\text{NMSD}(n)$. Moreover, there is no approach in the literature that approximately minimizes \eqref{eq:NMSDss} in heterogeneous networks, not even for the well-studied case of diffusion LMS networks with different step sizes in the nodes.}

{In this work, we follow a different approach. We propose an update rule driven by the online minimization of network MSE (NMSE), defined as
\begin{equation}
\label{network_MSE}
\text{NMSE}(n)=\frac{1}{N}\sum_{k=1}^N{\text{MSE}_k(n)}=\frac{1}{N}\sum_{k=1}^N{\mathbb{E}\{\check{e}^2_k(n)\}},
\end{equation}
where $\check{e}_k(n) = d_k(n)-\check{y}_k(n)=d_k(n)-{\bf{u}}_k^{\T}(n){\bf{w}}_k(n-1)$ represents the error at node $k$ using combined estimates available at that node, while $\check{y}_k(n)$ stands for the corresponding combined output.}

{It is well-known that for linear regression problems both criteria, MSD and MSE, are tightly related \cite{sayed2008adaptive}. In our approach, $\text{MSE}_k(n)$ can be easily approximated during both the convergence and the steady-state phases, and whatever the kind of update implemented by each node of the network. In addition, other advantage of using the NMSE as the optimization criterion lies in the fact that its minimization can be tackled with well known algorithms to update the combination weights, including gradient-based or LS strategies. In this respect, while using the NMSD requires a model for the network performance to overcome the lack of knowledge of the optimum weight vector, minimization of the network MSE can be seen as a model-free approach.}

\begin{figure}[t]
	\centering
	\includegraphics[width=0.8\linewidth]{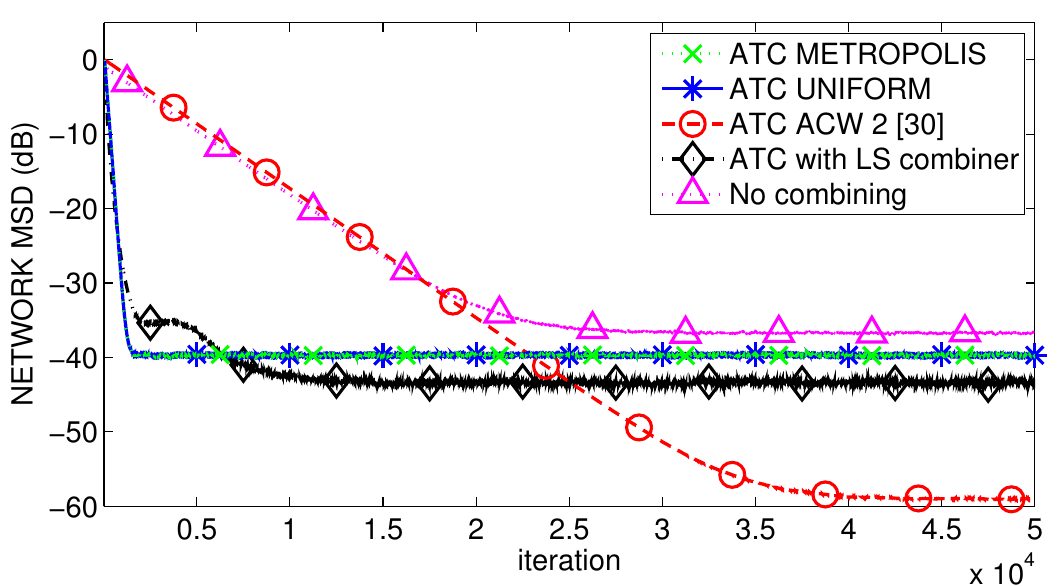}
	\caption{Estimation error for an ATC network with different combiners. It can be observed that state-of-the-art adaptive combination rules (in red) obtain a very good steady-state behaviour, much better that the ATC with LS combiners (black) because of the numerical problems of the latter.}
	\label{fig:atc_lse}
\end{figure}

{Fig. \ref{fig:atc_lse} illustrates the performance of a standard ATC diffusion in a fully connected network with 5 LMS nodes. As in the example in Fig. \ref{fig:ATCdiffusion_static}, one node has a comparably larger step size than the other nodes. The scheme of \cite{tu2011optimal}, with adaptive combination weights (ATC ACW 2), is not able to exploit the fast convergence of the node with large step size. This is due to the lack of an accurate NMSD approximation during convergence. When the network combiners are adjusted using an LS criterion (ATC with LS combiner), the behavior of the network is similar to the use of static combiners, what made us wonder why the network did not fully exploit the best properties of the different nodes. When analyzing this problem in detail, we observed severe numerical problems due to the coupled nature of adaptation and combination stages in standard ATC, i.e., the feedback of the combined weights in the adaptation step causes almost perfect correlation among the weight estimations at all nodes, ill-conditioning the minimization of \eqref{network_MSE}.}


{In the next section, we propose a novel diffusion scheme that overcomes these numerical instabilities, permitting an effective use of NMSE-based adaptive combiners.}



\section{Decoupled adapt-then-combine diffusion}
\label{sec:datc}

{Recently, we proposed a diffusion method \cite{fernandez2015distributed,fernandez2012novel} that also iterates an adaptation and a combination phase.} However, differently from standard ATC or CTA diffusion \cite{sayed2012diffusion,Sayed_SPM2013}, each node in our scheme preserves and adapts a purely local estimation $\psiM_k(n)$, which is then combined with the combined estimates, $\w_{\ell}(n-1)$, received from the neighboring nodes $\ell \in \mathcal{\bar{N}}_k$ at the previous iteration. Note that, although we have selected an ATC approach as the basis of our algorithm, it could be straightforwardly extended to CTA. Consequently, the proposed diffusion scheme can be written as follows
\begin{align}
\psiM_k(n)&= f_k ( \psiM_k(n-1), \uM_k(n) , d_k(n), \boldsymbol{\eta}_k), \label{eq:diffusion1}\\
\w_k(n)&= c_{kk}(n) \psiM_k(n) + \sum_{\ell\in\bar{\cal N}_k}c_{\ell k}(n)\w_{\ell}(n-1).\label{eq:diffusion2}
\end{align}
{with adaptive combiners $c_{\ell k}(n)$ selected to minimize \eqref{network_MSE}.}

In the adaptation phase \eqref{eq:diffusion1}, an updated local estimation $\psiM_k(n)$ is calculated as a function of the previous local estimation $\psiM_k(n-1)$, local data $\{d_k(n),\uM_k(n)\}$ and a state vector $\boldsymbol{\eta}_k$. In the combination phase \eqref{eq:diffusion2}, each node calculates $\w_k(n)$ using time-varying combination coefficients $c_{\ell k}(n)$.

{Two conditions are applied in the adaptation of $c_{\ell k}(n)$. Firstly, as most adaptive filtering schemes converge to unbiased estimations of the optimal solution in stationary scenarios, i.e., $\EE\{\wo(n)-\psiM_k(n)\} \rightarrow 0$ as $n \rightarrow \infty$, we constrain all coefficients at each node to sum up to one, in order to keep combined weights estimations unbiased in steady state.} In addition to this, to guarantee mean stability of our scheme, and in contrast with our previous work \cite{fernandez2015distributed}, we also impose non-negativity constraints on such combiners,
\begin{equation}
c_{\ell k}(n)\geq 0 \text{, } \sum_{\ell \in \mathcal{N}_k}c_{\ell k}(n)=1, \forall k.
\end{equation}
{These conditions on combination coefficients have also been considered in other diffusion schemes available in the literature to guarantee certain stability properties.}

{The adaptation and combination stages on the proposed diffusion algorithm can be interpreted in the following way: with respect to the adaptation of local estimates $\psiM_k(n)$, each node could be considered as an isolated adaptive filter working independently from the rest of the network.  Thus, it pursues the minimization of $\text{MSE}_k(n)$ using just its own regressors. During the combination phase, which includes the computation of combined weight estimators $\w_k(n)$ and the adaptation of the combiners, minimization of the network MSE is pursued.}

{We should emphasize that the most significant difference of \eqref{eq:diffusion1} and \eqref{eq:diffusion2} with respect to standard ATC is that the weight vector resulting from the combination is not fed back to the update of each individual filter. Even though under some circumstances this feedback can be beneficial, for instance to reduce the steady-state error in homogeneous networks, it also dilutes the differences between individual filter performances, which are key to get the maximum advantage out of heterogeneous networks. Furthermore, local estimates differ more among nodes than combined estimates, what results in a better conditioning for the model-free MSE-based adjustment of network combiners.}

{Since this diffusion scheme keeps the updates at each node decoupled from the rest of the network, we will refer to it in the following as {\em Decoupled ATC} (D-ATC). There are some additional advantages of decoupling the adaptation step from the combination phase:}

\begin{itemize}
\item Since nodes are updated as if they were working isolated from the network, the analysis of the local estimates can rely on existing models for adaptive filters. 

\item {Decoupled adaptation simplifies the design of heterogeneous networks, thus making it easy to include nodes that use different learning rules (e.g., LMS and RLS, Recursive Least-Squares), different learning parameters (e.g., step sizes, or asymmetry parameters in sparsity-aware nodes), or different filter lengths (using zero-padding for the shorter nodes).}

\item Related to this, the adaptation phase of our scheme is not influenced by an erroneous selection of the combination weights. In contrast, in standard ATC, if the combination weights are suboptimal, the adaptation phase of the diffusion algorithm is also affected.

\item Since the adaptation of each node is completely independent of other nodes' adaptation, we can more easily deal with synchronization issues. Furthermore, the combination stage can be modified to include the last available estimates received from the neighbors so that a delay in a particular node does not slow down the network.
\end{itemize}


%

\section{Theoretical Analysis of D-ATC}
\label{sec:performanceanalysis}

In this section, we analyze the performance of the D-ATC diffusion strategy in the mean and mean-square sense and
derive expressions for the steady-state NMSD in stationary and nonstationary environments.
Different from \cite{fernandez2015distributed,fernandez2013improved} and thanks to the energy conservation method \cite{sayed2008adaptive}, we directly obtain steady-state results, bypassing several of the
difficulties encountered when obtaining them as a limiting case of a transient analysis.
In order to simplify the analysis, the combiners $c_{\ell k}(n)$ are assumed to be static. Finally, {for this analysis we consider LMS and NLMS adaptations and consequently the general Equation \eqref{eq:diffusion1} becomes}
\begin{equation}
\psiM_k(n)=\psiM_k(n-1)+{\mu}_k(n) \uM_k(n) e_k(n), \label{eq:diffusionNLMS}
\end{equation}
where the local estimation error signals are
\begin{equation}
e_k(n)=d_k(n)-\uM_k^{T}(n) \psiM_k(n-1)\triangleq d_k(n)-y_k(n),
\end{equation}
{with $y_k(n)$ being the local output, and ${\mu}_k(n)$ a step size. For LMS, we have a constant step size ${\mu}_k(n)=\mu_k$, whereas for NLMS  ${\mu}_k(n)=\tilde{\mu}_k/\left[{\delta+\|\uM_k(n)\|^2}\right]$, with  $0<\tilde{\mu}_k<2$, with $\delta$ a regularization factor to prevent division by zero.}

\subsection{Data model and definitions}
\label{sec:data_model_definitions}

We start by introducing several assumptions to make the analysis more tractable. In order to obtain the most general results, during our analysis, we will delay the application of the different assumptions as much as possible.

\textbf{A1)} The unknown parameter vector $\wo(n)$ follows a \textit{random-walk model} \cite{sayed2008adaptive}. According to this widespread model, the optimal solution varies in a
nonstationary environment as
\begin{equation}
\wo(n)=\wo(n-1)+\q(n),\label{eq:rwm}
\end{equation}
where $\q(n)$ is a zero-mean, independent and identically distributed (i.i.d.) vector with autocorrelation matrix ${{\mathbf{Q}}}= \EE\{\q(n)\q^{\T}(n)\}$,
independent of the initial conditions $\psiM_k(0),\; \w_k(0)$, and of $\{\uM_k(n'), v_k(n')\}$
for all $k$ and $n'$. Although this model implies that the covariance matrix of $\wo(n)$ diverges as $n\to\infty$, it has been commonly used in the literature to keep the analysis of adaptive systems simpler \cite{sayed2008adaptive}. For ${{\mathbf{Q}}}= \mathbf{0}_M$ all expressions in the subsequent analysis are particularized for the stationary case.

\textbf{A2)} Input regressors are zero-mean
and have covariance matrix ${{{\mathbf{R}}}}_{k}=\EE\{\uM_k(n)\uM_{k}^{\T}(n)\}$.
Furthermore, they are spatially independent, i.e.,
\begin{equation}
\EE\{\uM_k(n)\uM_{\ell}^{\T}(n)\}= {\mathbf{0}}_M, \;\; k\neq\ell.\nonumber
\end{equation}
This assumption is widely employed in the analysis of diffusion algorithms and is realistic in many practical applications \cite{sayed2012diffusion}.
Furthermore, the noise processes $\{v_k(n)\}$
are assumed to be temporally white and spatially independent,
\begin{align}
\EE\{v_k(n)v_k(n')\}&=0,\;\; \text{for all}\;\;{n\neq n'},\nonumber\\
\EE\{v_k(n)v_{\ell}(n')\}&=0,\;\; \text{for all}\;\; n, n'\;\;\text{whenever}\;\; k\neq \ell.\nonumber
\end{align}
  Additionally, noise is assumed to be independent (not only uncorrelated) of the regression data $\uM_{\ell}(n')$, so that
$\E\{v_k(n)\uM_{\ell}(n')\}={\mathbf{0}}_M,\;\;\text{for all}\;\; k, \ell, n,\;\text{and}\; n'.$
{As a result, $\psiM_k(n-1)$ is independent of $v_{\ell}(n)$  for all $k$~and~$\ell$.
Since the regressors are assumed spatially independent,
$\psiM_k(n-1)$ is also independent of $\uM_{\ell}(n)$ for $k\neq \ell$.
For $k=\ell$, this independence condition also holds if the regressors are temporally uncorrelated.}
%

\textbf{A3)} {We will finally assume sufficiently small step sizes to neglect the effects of the statistical dependence of $\psiM_k(n-1)$ and $\uM_{k}(n)$ for colored regressors. This assumption has also been widely  used in analyses of diffusion schemes~\cite{sayed2012diffusion,Sayed_SPM2013,Takahashi2010,zhao2012performance, tu2011optimal}. Furthermore, results obtained from the independence
assumption between $\psiM_k(n-1)$ and $\uM_{k}(n)$ tend to match reasonably well the real filter performance for sufficiently small step sizes, even when
the temporal whiteness condition on the regression
data does not hold (see e.g., \cite{sayed2008adaptive}).}

To analyze adaptive diffusion strategies, it is usual to define weight-error vectors, taking into account the local and combined estimates of
each node, i.e.,
\begin{align}
\psitil_k(n)&\triangleq\wo(n)\!-\!\psiM_k(n),\\
\wtil_k(n)&\triangleq\wo(n)\!-\!\w_k(n),
\end{align}
with $k\!=\!1,\ldots,N$.

For notational convenience, we collect all weight-error vectors and products $v_k(n)\uM_k(n)$
across the network  into column vectors:
\begin{align}
&\!\!\!\wtilt(n)\!=\! \col\{\psitil_1(n),\cdots,\psitil_N(n),\wtil_1(n),\cdots,\wtil_N(n)\},\label{def:wtil}\\
&\!\!\!\boldsymbol{s}(n)\!\!=\!\col\{v_1(n)\uM_1(n), v_2(n)\uM_2(n),\cdots,v_N(n)\uM_N(n)\}\label{def:s},
\end{align}
where $\col\{\cdot\}$ represents the vector obtained by stacking its entries on top of each other. Note that the length of $\wtilt(n)$ is equal to $2MN$, whereas the length of $\boldsymbol{s}(n)$ is $MN$.
We also define the $(2MN)$-length column vector
\begin{equation}
\q_a(n)=\col\{\q(n), \q(n),\cdots\!, \q(n)\},\label{def:qa}
\end{equation}
and  the following $MN\times MN$ block-diagonal matrices containing the step sizes and information related to the autocorrelation matrices of the regressors:
\begin{align}
\Mcal(n)&= \diag\{{\mu}_1(n)\mathbf{I}_M,{\mu}_2(n)\mathbf{I}_M,\cdots,{\mu}_N(n)\mathbf{I}_M\},\label{def:Mcal}\\
\Rcal(n)&= \diag\{\uM_1(n)\uM_1^{\T}(n),\cdots,\uM_N(n)\uM_N^{\T}(n)\},\label{def:Rcal}
\end{align}
where
$\diag\{\cdot\}$ generates a block-diagonal matrix from its arguments and $\mathbf{I}_M$ is the $M\times M$ identity matrix. Finally, we also define the following matrices containing the combination weights:
\begin{align}
\mathbf{C}_{1}&= \diag\{c_{11},c_{22},\cdots,c_{NN}\},\label{def:C1}\\
\mathbf{C}_{2}&= \left[\begin{array}{cccc}
  0 & c_{12} & \cdots & c_{1N} \\
  c_{21} & 0 & \cdots & c_{2N} \\
  \vdots  & \vdots  & \ddots & \vdots  \\
  c_{N1} & c_{N2} & \cdots & 0   \end{array}\right],\label{def:C2}
\end{align}
and their extended versions
\begin{equation}
\Ccal_{i}\triangleq\mathbf{C}_{i}\otimes\mathbf{I}_M,\;\;\; i=1,2,\label{def:Ccal}
\end{equation}
 where $\otimes$ represents the Kronecker product of two matrices.

As a measure of performance, we consider the steady-state $\text{MSD}_k$ at each node and the steady-state NMSD, as defined in \eqref{eq:NMSDss}.

\subsection{Mean Stability Analysis}
\label{sec:meananalysis}

First, we present the mean convergence and stability analysis of our scheme. To do so, we start subtracting
 both sides of \eqref{eq:diffusion2} and \eqref{eq:diffusionNLMS} from $\wo(n)$. Under Assumption \textbf{A1}, using \eqref{eq:dk_model} and recalling that
$c_{kk}+\sum_{\ell\in\bar{\cal N}_k}c_{\ell k}=1$,
we obtain
\begin{align}
\psitil_k(n)\!-\!\q(n)&\!=\!\mathbf{A}_k(n)\psitil_k(n\!-\!1)\!-\!{\mu}_k(n)v_k(n)\uM_k(n), \label{eq:diffusion1til}\\
\wtil_k(n)\!-\!\q(n)&\!=\! c_{kk}\mathbf{A}_k(n) \psitil_k(n\!-\!1)\!+\!\!\!\sum_{\ell\in \bar{\cal N}_k}\!\!c_{\ell k}\wtil_{\ell}(n\!-\!1)
- c_{kk}{\mu}_k(n)v_k(n)\uM_k(n),\label{eq:diffusion2til}
\end{align}
 where
$\mathbf{A}_k(n)\triangleq\mathbf{I}_M-{\mu}_k(n)\uM_k(n)\uM_k^{\T}(n)$.

From \eqref{eq:diffusion1til} and \eqref{eq:diffusion2til}, using  definitions \eqref{def:wtil}-\eqref{def:Ccal}, and following algebraic manipulations similar to those of \cite{sayed2012diffusion}, we obtain the following equation characterizing the evolution of the weight-error vectors:
\begin{equation}
\wtilt(n) - \q_a(n) = \Bcal(n) \wtilt(n-1) - \boldsymbol{z}(n),
\label{eq:w_iteration}
\end{equation}
where
\begin{align}
\Bcal(n)& \triangleq  \left[\begin{array}{cc}
    \Bcal_{11}(n)&{\mathbf 0}_{(MN)}\\
    \Bcal_{21}(n)&  \Bcal_{22}
 \end{array}\right],\nonumber\\
 \Bcal_{11}(n)&=\mathbf{I}_{(MN)} - \Mcal(n)\Rcal(n),\nonumber\\
 \Bcal_{21}(n)&=\Ccal_{1}^{\T}[\mathbf{I}_{(MN)} - \Mcal(n)\Rcal(n)],\nonumber\\
  \Bcal_{22}&=\Ccal_{2}^{\T},\nonumber\\
 \boldsymbol{z}(n)& \triangleq[\Mcal(n)\boldsymbol{s}(n)\;\;\;\;\Ccal_{1}^{\T}\Mcal(n)\boldsymbol{s}(n)]^{\T}.\nonumber
 \end{align}

Under Assumptions \textbf{A2} and \textbf{A3}, all regressor vectors $\uM_k(n)$ are independent of $\psitil_{\ell}(n-1)$ and $\wtil_\ell(n-1)$ for $k,\,\ell=1,2,\cdots,N$.
Furthermore, independence of the noise w.r.t. the rest of variables implies that $\mathbb{E}\{\boldsymbol{s}(n)\} = \mathbf{0}_M$ and $\mathbb{E}\{\boldsymbol{z}(n)\}=\mathbf{0}_M$.
Thus, taking expectations on both sides of \eqref{eq:w_iteration} and recalling that $\mathbb{E}\{\q_a(n)\}=\mathbf{0}_{2MN}$, we obtain
  \begin{equation}
  \mathbb{E}\{\wtilt(n)\}=\mathbb{E}\{\Bcal(n)\}\mathbb{E}\{\wtilt(n-1)\}.
  \label{eq:Ewtil21}
  \end{equation}

A necessary and sufficient condition for the mean stability of \eqref{eq:Ewtil21} is that the
spectral radius of $\mathbb{E}\{\Bcal(n)\}$ is {less than one}, i.e., {$$\rho(\mathbb{E}\{\Bcal(n)\})=\max_i\{\lambda_i\}< 1,$$} where $\rho(\cdot)$ denotes the spectral radius of its matrix argument and $\lambda_i$, with $i=1, 2, \cdots, 2MN$, are the eigenvalues of $\mathbb{E}\{\Bcal(n)\}$ \cite{sayed2012diffusion}. Since $\mathbb{E}\{\Bcal(n)\}$ is a block-triangular matrix, its eigenvalues are the eigenvalues of the blocks of its main diagonal, i.e., the eigenvalues of $\mathbb{E}\{\Bcal_{11}(n)\}$ and $\mathbb{E}\{\Bcal_{22}\}$ \cite{horn2012matrix}.

Focusing first on matrix $\mathbb{E}\{\Bcal_{11}(n)\}$, we notice that it is also a block-diagonal matrix, so the step sizes need to be selected to guarantee
  \begin{equation}
  \rho(\mathbb{E}\{\Bcal_{11}(n)\})=\max_{1\leq k \leq N} \rho\left(\mathbf{I}_M-{\overline{\mathbf{R}}}_k\right)< 1, \label{eq:conditionB11}
  \end{equation}
  where
  \begin{align}
  {\overline{\mathbf{R}}_k}&{\triangleq \EE\left\{{\mu}_k(n)\uM_k(n)\uM_k^{\T}(n)\right\}.}\\
  \intertext{{For LMS, this matrix reduces to}}
  {\overline{\mathbf{R}}_k}&{= {\mu}_k\EE\left\{\uM_k(n)\uM_k^{\T}(n)\right\}=\mu_k\mathbf{R}_k}\label{eq:RbLMS}\\
  \intertext{{and for NLMS, we have}}
  \overline{\mathbf{R}}_k&={\tilde{\mu}_k}\EE\left\{\frac{\uM_k(n)\uM_k^{\T}(n)}{\delta+\|\uM_k(n)\|^2}\right\}\label{eq:RbNLMS}.
  \end{align}
{Condition  \eqref{eq:conditionB11} will be ensured for LMS if the step sizes $\mu_k$ satisfy~\cite{sayed2008adaptive}
 \begin{equation}
 0 < \mu_k < \frac{2}{\lambda_{\max}(\mathbf{R}_k)},\;\;\; \text{for}\;\;k=1,2,\cdots, N, \label{eq:conditionB11_LMS}
  \end{equation}
in terms of the largest eigenvalue of $\mathbf{R}_k$. Similarly, Condition \eqref{eq:conditionB11} will be ensured for NLMS if the step sizes $\tilde{\mu}_k$ satisfy~\cite{sayed2008adaptive}}
 \begin{equation}
 0 < {\tilde{\mu}}_k < 2,\;\;\; \text{for}\;\;k=1,2,\cdots, N. \label{eq:conditionB11_2}
  \end{equation}
 {Conditions \eqref{eq:conditionB11_LMS} and \eqref{eq:conditionB11_2}, which are well-known results for the LMS and NLMS algorithms, respectively,} guarantee that
  the local estimators $\{\psiM_k(n)\}$ are asymptotically unbiased, i.e., $\mathbb{E}\{\psitil_k(n)\}\rightarrow \mathbf{0}_M$ as $n\rightarrow \infty$ for all nodes of the network.

For the spectral radius of ${\Bcal}_{22} = \Ccal_{2}^{\T}$, we can rely on the following bound from \cite{horn2012matrix}:
\begin{equation}
\rho(\Bcal_{22})\leq\|\Bcal_{22}\|_{\infty} =\displaystyle \max_k \sum_{\ell \in \bar{\mathcal{N}}_k} |c_{\ell k}|.\label{eq:conditionB22}
\end{equation}
A sufficient (but not necessary) condition to guarantee $\rho(\Bcal_{22})\leq 1$ is to keep all combination weights non-negative. In effect, since the sum of all combiners associated to a node is one, using non-negative weights we have
\begin{equation}
\label{conv_eq}
\rho(\Bcal_{22})\leq \displaystyle \max_k \sum_{\ell \in \bar{\mathcal{N}}_k} c_{\ell k} = \displaystyle \max_k\, (1 - c_{kk}) \leq 1.
\end{equation}

When combiners are learned by the network, non-negativity constraints can be applied at every iteration to ensure mean stability. Although our derivations show that this is just a sufficient condition, we should mention that in our previous simulation work of \cite{fernandez2015distributed, fernandez2012novel}, where we allowed combination weights to become negative, the network showed some instability problems {that have been removed thanks to the application of these constraints.}

\subsection{Mean-Square Performance}
\label{sec:meansquareanalysis}

We present next a mean-square performance analysis, following the energy conservation framework of \cite{sayed2008adaptive}. First, let $\Sgm$ denote an arbitrary nonnegative definite $2MN\times 2MN$ matrix. Different choices of $\Sgm$ allow us to obtain different performance measurements of the network \cite{Sayed_SPM2013}.

Thus, computing the weighted squared norm on both sides of \eqref{eq:w_iteration} using $\Sgm$ as
 a weighting matrix, we arrive~at
  \begin{align}
&\wtilt^{\T}\!(n)\Sgm\wtilt(n)\!-\!\wtilt^{\T}\!(n)\Sgm\q_a(n)\!-\!\q_a^{\T}\!(n)\Sgm\wtilt(n)\!+\!\q_a^{\T}\!(n)\Sgm\q_a(n)\nonumber\\
&\;\;\;\;\;\;=\wtilt^{\T}\!(n-1) \Bcal^{\T}\!(n)\Sgm \Bcal(n)\wtilt(n\!-\!1)+\!\boldsymbol{z}^{\T}\!(n)\Sgm\boldsymbol{z}(n)-\!2\boldsymbol{z}^{\T}\!(n)\Sgm\Bcal(n)\wtilt(n\!-\!1).
\label{eq:w_iterationsquare}
\end{align}

As before, independence of the noise terms in $\boldsymbol{z}(n)$ with respect to all other variables implies that the last element in \eqref{eq:w_iterationsquare} vanishes under expectation. Furthermore,
under Assumption \textbf{A1}, we can verify that
\begin{align}
\EE\{\wtilt^{\T}\!(n)\Sgm\q_a(n)\}&=\!\EE\{\q_a^{\T}\!(n)\Sgm\wtilt(n)\}
\!=\!\EE\{\q_a^{\T}\!(n)\Sgm\q_a(n)\}=\Tr(\Sgm\mathbf{Q}_a),\label{eq:qSq}
\end{align}
where  $\Tr(\cdot)$ stands for the trace of a matrix and
$$\mathbf{Q}_a\triangleq\EE\{\q_a(n)\q_a^{\T}(n)\}=\mathbf{J}_{(2N)}\otimes\mathbf{Q},$$
being $\mathbf{J}_{(2N)}$ a $2N\times 2N$ matrix with all entries equal to one. Defining the matrices
\begin{align}
\Scal&\triangleq {\diag\Big\{\sigma_{v_1}^2\EE\{{\mu}_1^2(n)\uM_1(n)\uM_1^{\T}(n)\},}
 {\sigma_{v_2}^2\EE\{{\mu}_2^2(n)\uM_2(n)\uM_2^{\T}(n)\},\;\;\cdots,}\nonumber \\
&\hspace*{1.5cm} {\sigma_{v_N}^2\EE\{{\mu}_N^2(n)\uM_N(n)\uM_N^{\T}(n)\}\Big\}},\label{def:Scal}\\
\Zcal&\triangleq \EE\{\boldsymbol{z}(n)\boldsymbol{z}^{\T}\!(n)\}=\left[\begin{array}{cc}
                                                                           \Scal & \Scal\Ccal_1 \\
                                                                           \Ccal_1^{\T}\Scal &  \Ccal_1^{\T}\Scal\Ccal_1
                                                                         \end{array}\right],\label{eq:Zcal}
\end{align}
using \eqref{eq:qSq}, and taking  expectations of both sides of \eqref{eq:w_iterationsquare}, we obtain
\begin{align}
\EE\{\|\wtilt(n)\|^2_{\Sgm}\}= &\;\EE \left\{\|\wtilt(n\!-\!1)\|^2_{{\Bcal}^{\T}(n)\Sgm {\Bcal}(n)}\right\} +\!\Tr(\Sgm\Zcal)\!+\!\Tr(\Sgm\mathbf{Q}_a),
\label{eq:w_iterationsquare2}
\end{align}
where  $\|\mathbf{x}\|^2_{\Sgm}$ denotes the weighted squared norm $\mathbf{x}^{\T}\Sgm\mathbf{x}$.

Using Assumption {\bf A3}, we can replace the random matrix $\Bcal(n)$ by its steady-state mean
value $\overline{\Bcal}=\lim_{n\rightarrow \infty}\EE\{\Bcal(n)\}$, which is equivalent to replacing the matrix
{$\mu_k(n)\uM_k(n)\uM_k^{\T}(n)$ by its mean $\overline{\mathbf{R}}_k$, given by \eqref{eq:RbLMS} for LMS or by \eqref{eq:RbNLMS} for NLMS.}
Using this approximation, \eqref{eq:w_iterationsquare2} reduces~to
\begin{align}
\EE\{\|\wtilt(n)\|^2_{\Sgm}\} \approx\; &\EE\left\{\{\|\wtilt(n\!-\!1)\|^2_{\overline{\Bcal}^{\T}\Sgm \overline{\Bcal}}\right\}+ \Tr(\Sgm\Zcal)\!+\!\Tr(\Sgm\mathbf{Q}_a).
\label{eq:w_iterationsquare3}
\end{align}

\subsubsection*{Mean-Square Convergence}

{As in \cite{Sayed_SPM2013},  the convergence rate of the series is governed by $[\rho(\overline{\Bcal})]^2$, in
terms of the spectral radius of $\overline{\Bcal}$.
From Section~\ref{sec:meananalysis}, we can obtain a superior limit for $\rho(\overline{\Bcal})$, which is given by
%
\begin{equation}\hspace{-0.1em}
\rho(\overline{\Bcal})\leq \max\left\{\max_{k,i} \left[1-{\lambda_i({\overline{\mathbf{R}}}_k)}\right],\;\;\max_{k}\left(1-c_{kk}\right)\right\}.\hspace{-1.5em}\label{eq:suplim}
\end{equation}
{Choosing the step size of the LMS (resp., NLMS) algorithm into the interval \eqref{eq:conditionB11_LMS} [resp., \eqref{eq:conditionB11_2}]} and imposing non-negativity constraints to the combiners, $\rho(\overline{\Bcal})\leq 1$ and the convergence of $\lim_{n\rightarrow \infty}\EE\{\|\wtilt(n)\|^2_{\Sgm}\}$ is ensured. Furthermore,
from the superior limit \eqref{eq:suplim}, we can see that, in the worst case,  our diffusion scheme can converge with the same convergence rate of the noncooperative solution, whose spectral radius is $\max_{k,i}\left\{1-{\lambda_i({\overline{\mathbf{R}}}_k)}\right\}$ (considering that all the nodes are adapted using {LMS or NLMS}). However, we show by means of simulations that in practice this limit is very conservative and the proposed diffusion scheme converges much faster than the noncooperative solution.}

\subsubsection*{Steady-state MSD performance}

It it important to notice that variance relations similar to \eqref{eq:w_iterationsquare3} have often appeared in the performance analysis of diffusion schemes \cite{Sayed_SPM2013}.
Iterating  \eqref{eq:w_iterationsquare3} and taking the limit as $n\rightarrow \infty$, we conclude that (see, e.g., \cite{Tu_TSP2013})
\begin{equation}
\lim_{n\rightarrow \infty}\EE\{\|\wtilt(n)\|^2_{\Sgm}\} \approx \sum_{j=0}^{\infty}\Tr[\overline{\Bcal}^j(\Zcal+\mathbf{Q}_a)(\overline{\Bcal}^{\T})^{j}\Sgm].
\label{eq:wtilseries}
\end{equation}
To obtain analytical expressions for the steady-state MSD
of the network and of its individual nodes, we will replace $\Sgm$ by the following matrices
\begin{align}
\boldsymbol{\Gamma}&\triangleq\left[\begin{array}{cc}
                            \mathbf{0}_{NM} & \mathbf{0}_{NM} \\
                             \mathbf{0}_{NM} & \frac{1}{N}\mathbf{I}_{NM}
                          \end{array}
\right],\label{eq:Gamma}\\
\boldsymbol{\Upsilon}_k&\triangleq\left[\begin{array}{cc}
                            \mathbf{0}_{NM} & \mathbf{0}_{NM} \\
                             \mathbf{0}_{NM} & \mathbf{E}_k\otimes\mathbf{I}_{M}
                          \end{array}
\right] ,\label{eq:Sgm}
\end{align}
where $\mathbf{E}_k$ is an $N\times N$ zero matrix, except in the element $(k,k)$, that is equal to one. Replacing $\Sgm$ in \eqref{eq:wtilseries} by either $\boldsymbol{\Gamma}$ or $\boldsymbol{\Upsilon}_k$, the MSD performance of the network and of its individual nodes can be expressed, respectively, by
\begin{align}
{\rm NMSD}(\infty)&\approx\sum_{j=0}^{\infty}\Tr[\overline{\Bcal}^j(\Zcal+\mathbf{Q}_a)(\overline{\Bcal}^{\T})^{j}\boldsymbol{\Gamma}], \label{eq:MSDnet1}\\
{\rm MSD}_{k}(\infty)&\approx\sum_{j=0}^{\infty}\Tr[\overline{\Bcal}^j(\Zcal+\mathbf{Q}_a)(\overline{\Bcal}^{\T})^{j}\boldsymbol{\Upsilon}_k].\label{eq:MSDnode1}
 \end{align}
Since $\overline{\Bcal}$ is lower triangular, matrix $\overline{\Bcal}^j$ is given by
\begin{equation}
\overline{\Bcal}^j=\left[\begin{array}{cc}
                      \overline{\Bcal}_{11}^j & \mathbf{0}_{(MN)} \\
                      \bar{\Xcal}(j) & \overline{\Bcal}_{22}^j
                    \end{array}
\right],\label{eq:Bj}
\end{equation}
being
\begin{align}
&\bar{\Xcal}(j)\!=\!\!\sum_{k=0}^{j-1}\overline{\Bcal}_{22}^k\overline{\Bcal}_{21}\overline{\Bcal}_{11}^{j\!-\!k\!-\!1}
\!=\!\!\sum_{k=0}^{j-1}{[\Ccal_2^{\T}]^k\Ccal_1^{\T}\left[\mathbf{I}_{(MN)}\!-\!\Lcal\right]^{j\!-\!k}},
\label{eq:Xcal}\\
\intertext{where we have defined}
&\Lcal\!\triangleq\!\lim_{n\rightarrow\infty}\!\!\EE\{\Mcal(n)\Rcal(n)\}\!=\!\diag\{\overline{\mathbf{R}}_1,\,\overline{\mathbf{R}}_2,\cdots,\overline{\mathbf{R}}_N\}.\label{eq:McalRcal}
\end{align}
Replacing \eqref{eq:Bj} and \eqref{eq:Zcal} respectively in \eqref{eq:MSDnet1} and \eqref{eq:MSDnode1}, we arrive at
\begin{align}
{\rm NMSD}(\infty)&\approx \frac{1}{N}\sum_{j=0}^{\infty}\Tr\bigg[\bar{\Xcal}(j)(\Scal\!+\!{\Qcal})\bar{\Xcal}^{\T}\!(j)
+2(\Ccal_2^{\T})^j(\Ccal_1^{\T}\Scal\!+\!{\Qcal})\bar{\Xcal}^{\T}\!(j)
+(\Ccal_2^{\T})^{j}(\Ccal_1^{\T}\Scal\Ccal_1\!+\!{\Qcal})\Ccal_2^j\;\;\bigg],\label{eq:MSDnet2}\\
{\rm MSD}_{k}(\infty)&\approx \sum_{j=0}^{\infty}\Tr\bigg[\Big(\bar{\Xcal}(j)(\Scal\!+\!{\Qcal})\bar{\Xcal}^{\T}\!(j)
+\!2(\Ccal_2^{\T})^j(\Ccal_1^{\T}\Scal\!+\!{\Qcal})\bar{\Xcal}^{\T}\!(j)
+\!(\Ccal_2^{\T})^{j}(\Ccal_1^{\T}\Scal\Ccal_1\!+\!{\Qcal})\Ccal_2^j\Big)\mathbf{E}_k\otimes\mathbf{I}_{M}\bigg],\label{eq:MSDnode2}
\end{align}
where $\Qcal=\mathbf{J}_{N}\otimes\mathbf{Q}$. Note that the $MN\times MN$ matrix $\Qcal$ is similar to matrix $\mathbf{Q}_a$, but has half its size.

{If all the nodes of the network update their local estimates with the LMS algorithm,
the theoretical steady-state MSD can be estimated by \eqref{eq:MSDnet2} and \eqref{eq:MSDnode2}, recalling
that matrices $\overline{\mathbf{R}}_k$, which appear in $\bar{\Xcal}(j)$,
are given by \eqref{eq:RbLMS} and that matrix $\Scal$ reduces to
\begin{equation}
\Scal=\diag\left\{\sigma_{v_1}^2\mu_1^2{\mathbf{R}}_1,\;\sigma_{v_2}^2\mu_2^2{\mathbf{R}}_2,\;\cdots,\; \sigma_{v_N}^2\mu_N^2{\mathbf{R}}_N\right\}.\label{eq:SLMS}
\end{equation}
On the other hand, assuming  NLMS adaptation,
we still have to obtain approximations for matrices $\overline{\mathbf{R}}_k$ and $\EE\{\mu_k^2(n)\uM_k(n)\uM_k^{\T}(n)\}$. For this purpose,
we assume:}

\textbf{A4) }
The number of coefficients $M$ is large enough for each element of the matrix $\uM_k(n)\uM_k^{\T}(n)$ to be approximately
independent from  $\sum_{l=0}^{M-1}|u(n-l)|^2$.
This is equivalent to applying the averaging principle of \cite{Samson1983}, since for large $M$, $\|\uM_k(n)\|^2$
tends to vary slowly compared to the individual entries of $\uM_k(n)\uM_k^{\T}(n)$.

\textbf{A5) }
The regressors $\uM_k(n)$, $k=1,2,\ldots, N$ are formed by a tapped-delay line with Gaussian entries and the regularization factor is equal to zero $(\delta=0)$. This is a common assumption in the analysis of adaptive filters and leads to reasonable analytical results
\cite{Haykin2002}.
Under \textbf{A4} and \textbf{A5}, we obtain the following approximations from \cite{Costa2002}:
\begin{align}
\overline{\mathbf{R}}_k&\approx {{\tilde{\mu}}_k}\frac{\mathbf{R}_k}{\sigma_{u_k}^2(M-2)},\label{approxRbark}\\
{\EE\{\mu_k^2(n)\uM_k(n)\uM_k^{\T}(n)\}}&\approx {{\tilde{\mu}}_k^2}\frac{\mathbf{R}_k}{\sigma_{u_{k}}^4(M-2)(M-4)}.\label{approxRtilk}
\end{align}
The model to compute the steady-state MSD of the network and of its individual nodes can be summarized as follows:
(i) compute the  matrices of the combination weights using \eqref{def:C1}-\eqref{def:Ccal} and the matrix $\Qcal$, according to the environment variation;
(ii) {for LMS (resp., NLMS) adaptation, use \eqref{eq:RbLMS} [resp., \eqref{approxRbark} and \eqref{approxRtilk}]}
in the computation of matrices  $\Scal$ and $\bar{\Xcal}(j)$, defined respectively by \eqref{def:Scal} and \eqref{eq:Xcal};
and finally, (iii) use these matrices in  \eqref{eq:MSDnet2} and \eqref{eq:MSDnode2}.

\section{NMSE-based adaptive combiners}
\label{sec:Adaptivecombiners}

As shown in Section \ref{sec:intro}, the implementation of adaptive combiners is crucial for {heterogeneous networks. For instance, when the nodes have different step sizes in the adaptation step, the combiners should favor the diffusion of the estimates of the fastest nodes during network convergence. However, the network should favor the nodes with better SNR and smaller adaptation step size in steady state, as they produce lower steady-state
misadjustment.}

{In this section we present two strategies for learning the combiners suitable for our Decoupled ATC scheme. These two strategies are based on an approximate minimization of the network Mean-Square Error at each step $n$ as defined in \eqref{network_MSE}. Since every node only optimizes its own combination coefficients, this is equivalent to minimizing $\text{MSE}_k(n)$ node-wise. As stated in Subsection \ref{sec:adaptive_combiners_hetero}, there are different well-known algorithms that can be used to optimize $\text{MSE}_k(n)$ including gradient-based or LS strategies. However, it should be remarked that due to the nature of the problem, in particular because of the expected large correlation among the solution estimates shared by the nodes, not all adaptive algorithms to update $c_{\ell k}(n)$ would obtain a competitive performance. In this work, we include two approaches to adapt the combination coefficients that have demonstrated their benefits with respect to other schemes.}






Finally, let us recall that we would like to satisfy the convexity constraint of Subsection \ref{sec:meananalysis} to guarantee stability (and also to follow the criterion of other works in this field, e.g., \cite{sayed2012diffusion, Takahashi2010, zhao2012performance, cattivelli2008diffusion, tu2011optimal}). Since a direct application of the algorithms below may give rise to values of $c_{\ell k}(n)$ outside range $[0,1]$, we will enforce the combination parameters $c_{\ell k}(n)$ to remain in the desired interval $[0,1]$ at each iteration. {For simplicity,} if any $c_{\ell k}(n)$ results negative after its update, we simply set it to zero and then rescale the remaining combination weights so that they sum up to one. {We would like to remark that more complex projection rules could have been used to implement this constraint but, as the proposed method shows a good performance, we leave as future work the analysis and evaluation of alternative solutions.}

\subsection{Affine Projection Algorithm}
\label{sec:APA}

In this subsection we present an Affine Projection Algorithm (APA) for the stochastic minimization of the MSE in \eqref{network_MSE}.

{First, it is useful to define some notation.} We stack the combination coefficients $c_{\ell k}(n)$ of node $k$, with $\ell \in \bar{\cal N}_k$, in a length-$\bar{N}_k$ vector ${\bf{\bar{c}}}_{k}(n)$. Doing so, we can write
\begin{equation}
\label{eq:c_kk}
c_{k k}(n) = 1- \sum_{\ell\in\bar{\cal N}_k}c_{\ell k}(n)=1-{\mathds{1}}_{\bar{\cal N}_k}^{\T}{\bf{\bar{c}}}_{k}(n).
\end{equation}
Then, defining $y_{\ell k}(n)=\uM_k^{\T}(n)\w_{\ell}(n-1)$ and $\tilde{y}_{\ell k}(n)=y_{\ell k}(n)-y_k(n)$ with $\ell \in \bar{\cal N}_k$, collecting all these differences into a column vector $\tilde{\mathbf{y}}_k(n)$, and using \eqref{eq:c_kk}, ${\rm MSE}_k(n)$ can be rewritten as
\begin{equation}
\label{MSE_apa}
\text{MSE}_k(n) = \mathbb{E}\left\{ \left[e_k(n)-\bar{\bf{c}}_k^{\T}(n)\tilde{\mathbf{y}}_k(n)\right]^2\right\}.
\end{equation}
Applying the standard APA algorithm \cite{sayed2008adaptive} to minimize this cost function, we obtain a regularized affine projection algorithm for the adaptation of ${\bf{\bar{c}}}_{k}(n)$:

\begin{equation}
\label{APA}
{\bf{\bar{c}}}_{k}(n) ={\bf{\bar{c}}}_{k}(n-1) + \mu_c[\epsilon {\bf{I}}_{\bar{N}_k}+{\bf{\tilde{Y}}}_{k}^{\T}(n){\bf{\tilde{Y}}}_{k}(n)]^{-1} {\bf{\tilde{Y}}}^{\T}_{k}(n) \times 
[{\bf{e}}_k(n)-{\bf{\tilde{Y}}}_{k}(n){\bf{\bar{c}}}_{k}(n-1)],
\end{equation}
where $\mu_c$ is a step size to control the adaptation of ${\bf{\bar{c}}}_{k}(n)$, $\epsilon$ is a small regularization parameter to prevent division by zero, ${\bf{\tilde{Y}}}_{k}(n)$ is an $L\times\bar{N}_k$ matrix whose $L$ rows corresponds with the last $L$ values of vector $\tilde{\bf{y}}_k(n)$, ${\bf{e}}_k(n)=[e_k(n),e_k(n-1),\cdots,e_k(n-L+1)]^{\T}$, and ${\bf{I}}_{\bar{N}_k}$ represents the ${\bar{N}_k}\times {\bar{N}_k}$ identity matrix, with ${\bar{N}_k}$ the cardinal of $\bar{\mathcal{N}}_k$.

This recursion requires the inversion of an $\bar{N}_k\times\bar{N}_k$ matrix at each iteration, resulting in an attractive implementation if the projection order $L$ is larger than the number of neighbors of node $k$, $\bar{N}_k$. Otherwise, if for any node $\bar{N}_k>L$, we can invoke the matrix inversion lemma \cite{sayed2008adaptive} to rewrite \eqref{APA} as
\begin{equation}
\label{APA2}
{\bf{\bar{c}}}_{k}(n)={\bf{\bar{c}}}_{k}(n-1)+\mu_c{\bf{\tilde{Y}}}_{k}^{\T}(n)[\epsilon {\bf{I}}_L+{\bf{\tilde{Y}}}_{k}(n){\bf{\tilde{Y}}}^{\T}_{k}(n)]^{-1} \times [{\bf{e}}_k(n)-{\bf{\tilde{Y}}}_{k}(n){\bf{\bar{c}}}_{k}(n-1)],
\end{equation}
which requires the inversion of an $L \times L$ matrix. 

Equations \eqref{APA} --or \eqref{APA2}-- and \eqref{eq:c_kk}, constitute the $\epsilon$-APA algorithm for adapting the combiners at each node. More details about the derivation are provided in Appendix \ref{apen:APA}.

\subsection{Least-Squares Algorithm}
\label{sec:LS}

In this subsection, we follow a Least-Squares approach similar to the one in \cite{fernandez2015distributed} and \cite{fernandez2012novel}. Instead of minimizing \eqref{network_MSE} using a stochastic minimization algorithm, we replace ${\text{MSE}_k(n)}$ by the following related cost function \cite{sayed2008adaptive},
\begin{equation}
J_k(n)=\sum_{i=1}^{n}\beta(n,i)\check{e}_k^2(n,i),\label{eq:JkLS}
\end{equation}
where $\beta(n,i)$ is a temporal weighting window, and
\begin{equation}
\check{e}_k(n,i)=e_k(i)-\bar{\bf{c}}_k^{\T}(n)\tilde{\mathbf{y}}_k(i)
\label{eq:eni}
\end{equation}
represents the error incurred by node $k$ at time $i$ when the outputs of all nodes belonging to ${\cal N}_k$ are combined using the combiners at time
$n$.



%

Following a standard LS method to minimize this cost function, we obtain
\begin{equation}
\bar{\mathbf{c}}_k(n)= \left( \mathbf{P}_k(n) + \epsilon {\bf I}_{\bar{N}_k} \right)^{-1}\mathbf{z}_k(n),\label{eq:Pcz2}
\end{equation}
where a small regularization constant $\epsilon$ is again introduced since $\mathbf{P}_k(n)$ could be ill-conditioned \cite{fernandez2012novel}.
Similarly to the case of combination of multiple filters \cite{Azpicueta-Ruiz2010a}, $\mathbf{P}_k(n)$ can be interpreted as the autocorrelation matrix of vector $\tilde{\mathbf{y}}_k(n)$ while $\mathbf{z}_k(n)$ would be seen as the cross-correlation vector between $\tilde{\mathbf{y}}_k(n)$ and $e_k(n)$.

For further details about the derivation of the LS algorithm, please refer to Appendix \ref{apen:LS}.

\subsubsection*{Temporal weighting window}

{The temporal weighting window $\beta(n,i)$, in the cost function \eqref{eq:JkLS} and the computation of  $\mathbf{P}_k(n)$ and $\mathbf{z}_k(n)$, deserves some discussion.} In this paper, we propose the use of an exponential weighting window,
\begin{equation}
\beta(n,i) = \gamma^{n-i},
\end{equation}
where $\gamma$ is a forgetting factor $0 < \gamma \leq 1$.

This contrasts with our choice in previous works \cite{fernandez2015distributed,fernandez2012novel}, where we leaned towards a rectangular window, which provided a good convergence but a worse steady-state performance than standard ATC with adaptive combiners \cite{tu2011optimal}. The reason for that choice was the instability problems of affine combiners when long windows were used. In this paper, as we use the more stable convex combiners (see Subsection \ref{sec:meananalysis}) an exponential window can be safely employed. This window has two remarkable advantages with respect to a rectangular window: 1) It is more efficient in terms of memory and computation; and 2) it allows a recursive implementation. In addition, as we show in the experiments in the next section, the LS algorithm with exponential window outperforms other state-of-the art approaches.

\section{Simulation Results}
\label{sec:simulations}


In this section we present a number of simulation results to illustrate the behavior of D-ATC and the proposed adaptive combiners rules in stationary estimation and tracking scenarios. In the simulations, we consider only the NLMS algorithm to update the nodes due to its inherent advantages with respect to LMS. {Nevertheless, it should be remarked that we have carried out experiments where nodes are updated with the LMS algorithm obtaining similar conclusions.}

\begin{figure}[t]
	\begin{center}
		\begin{tabular}{cc}
			\includegraphics[width=0.38\linewidth]{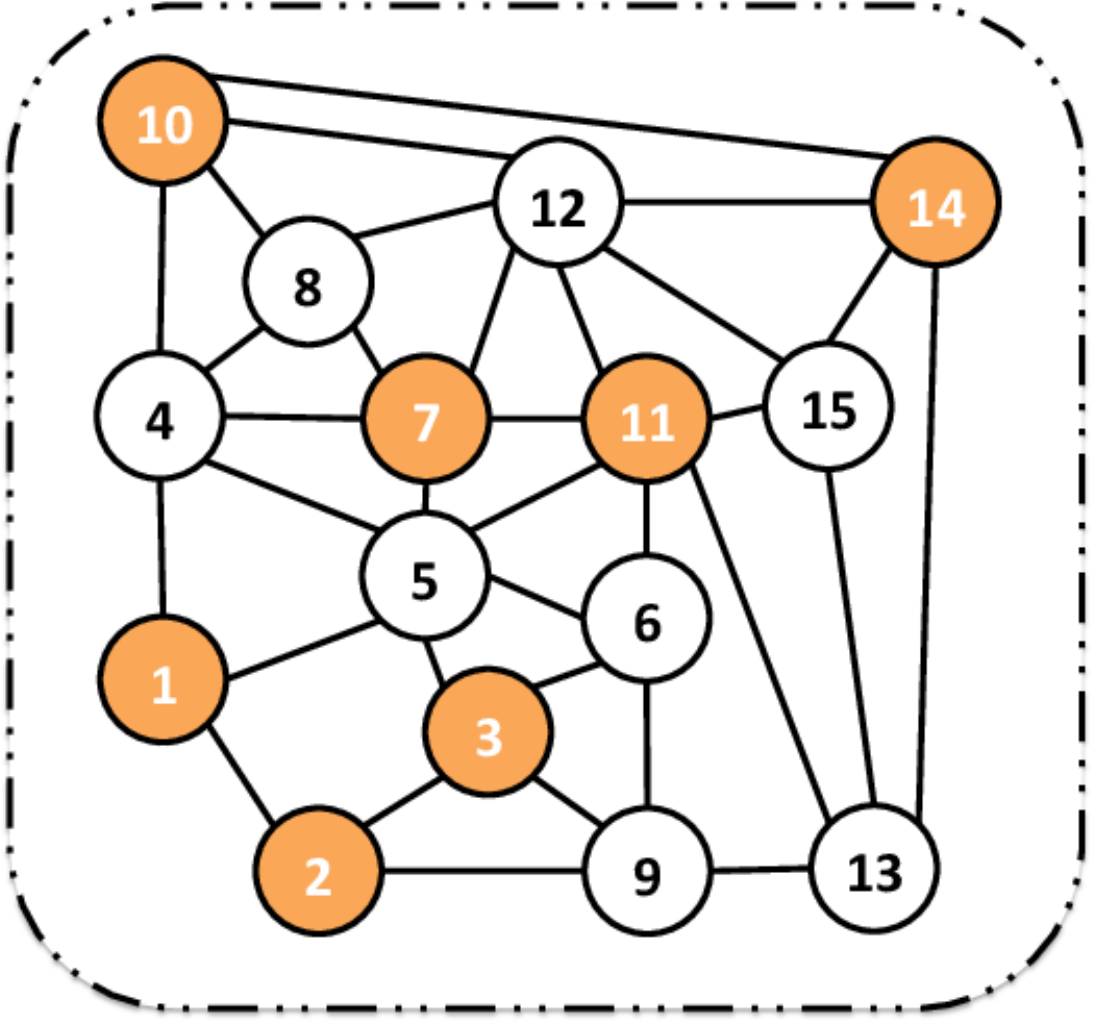}& \includegraphics[width=0.45\linewidth]{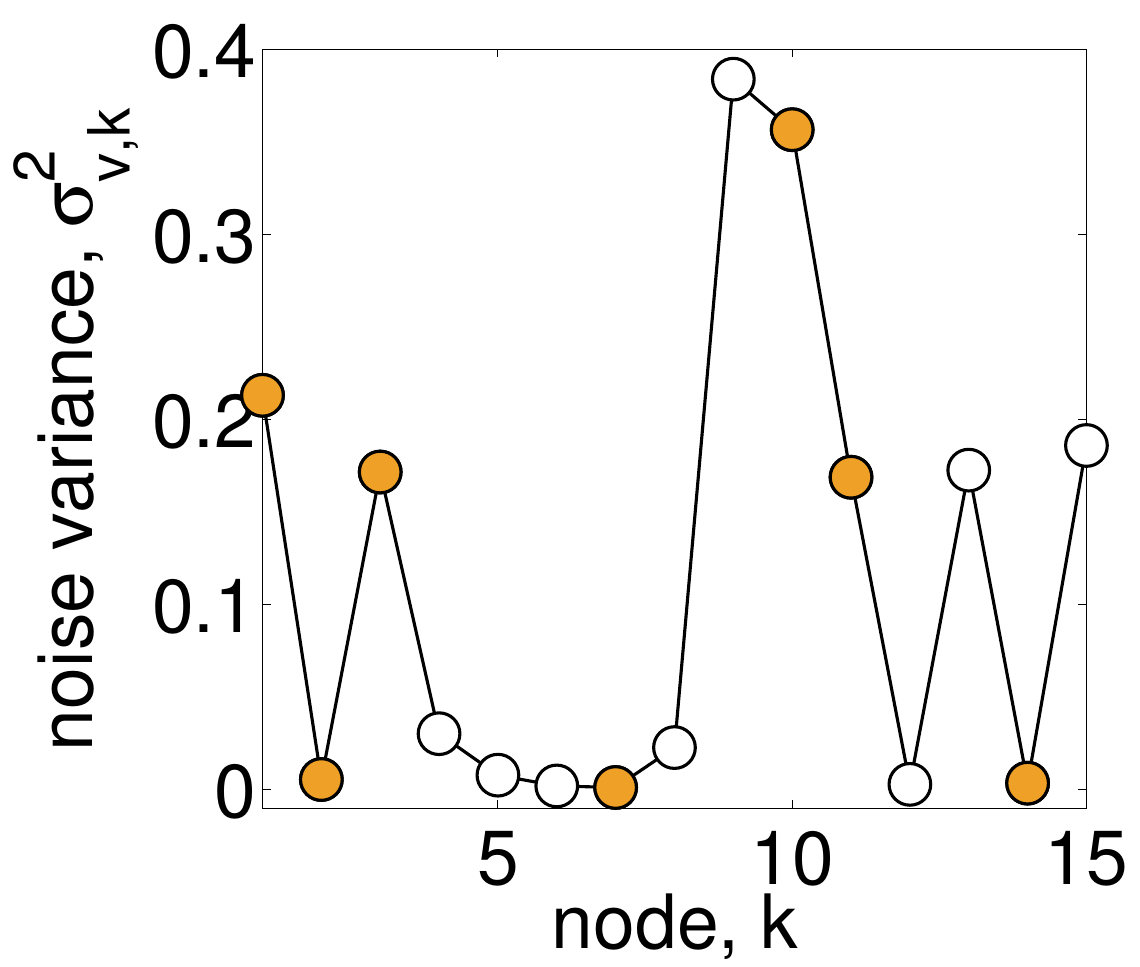}\\
			\footnotesize{(a)} & \footnotesize{(b)}
\vspace*{-0.2cm}
		\end{tabular}
		\caption{(a) Network topology for the simulation experiments: orange shaded nodes are adapted with $\tilde{\mu}_k=0.1$ and the rest with $\tilde{\mu}_k=1$. (b) noise power $\sigma^2_{v,k}$ at each node in the network.}
		\label{fig:network}
		\vspace*{-0.5cm}
	\end{center}
\end{figure}

We simulate the 15-node network of Fig. \ref{fig:network}(a), where all the nodes employ NLMS adaptation. The nodes step sizes are taken as $\mu_k \in \{0.1, 1\}$ as illustrated in Fig. \ref{fig:network}. The input signals $\uM_k(n)$ follow a multidimensional Gaussian with zero mean and the same scalar covariance matrix, $\sigma_{u}^2{\bf{I}}_M$, with $\sigma_{u}^2=1$. Unless otherwise stated, the observation noise $v_k(n)$ at each node is also Gaussian distributed with zero mean and variance $\sigma_{v,k}^2$ randomly chosen between $[0,0.4]$ as shown in Fig. \ref{fig:network}(b). For the stationary estimation problem, the parameter vector $\wo$ is a length-50 vector with values uniformly taken from range $\left[-1,1\right]$. As a tracking model, we use the one presented in \eqref{eq:rwm}.

First, we present a set of experiments with the aim to validate the theoretical analysis of Section \ref{sec:performanceanalysis}. Then, we compare the behavior of our rules to state-of-the-art adaptive combination algorithms for standard ATC \cite{tu2011optimal}, both in stationary and tracking scenarios.

\subsection{Validation of the theoretical analysis for D-ATC}
\label{sec:exp_validation}
\begin{figure}[t]
	\begin{center}
		\begin{tabular}{cc}
			\includegraphics[width=0.4\linewidth]{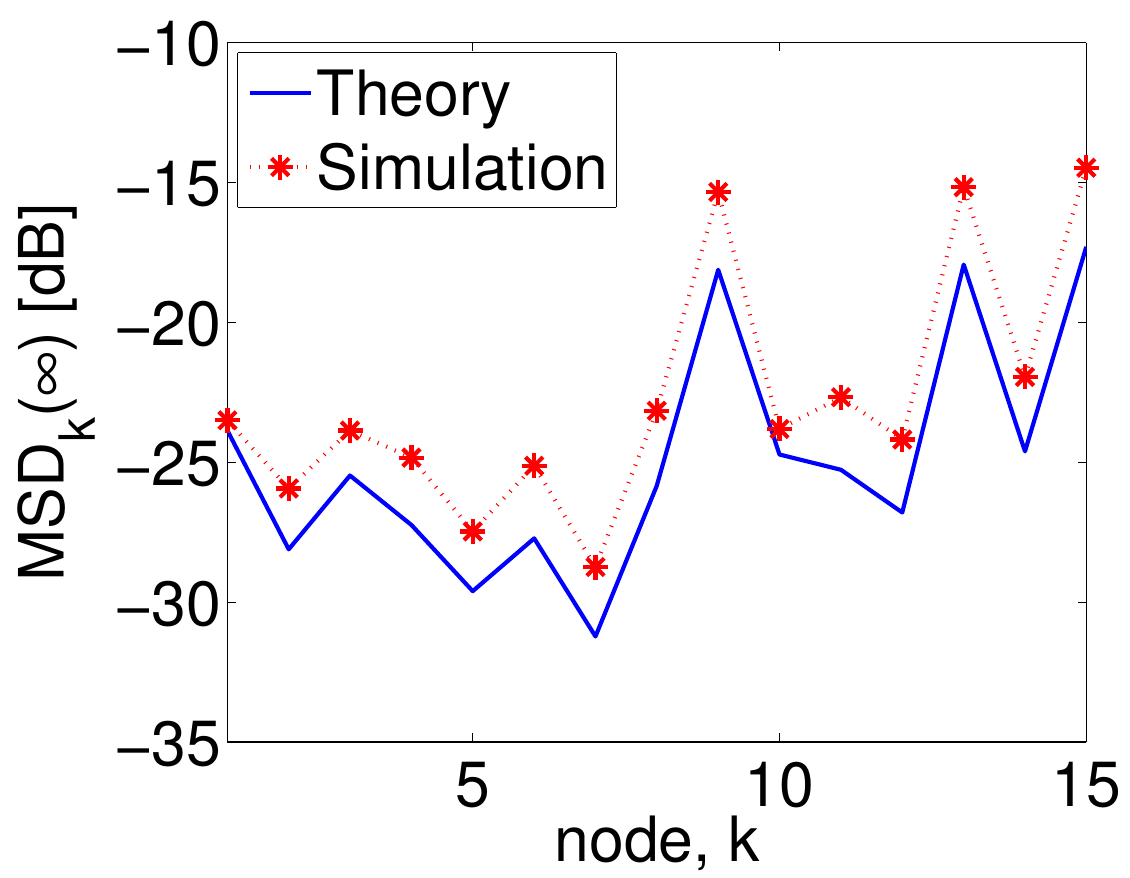} & \includegraphics[width=0.4\linewidth]{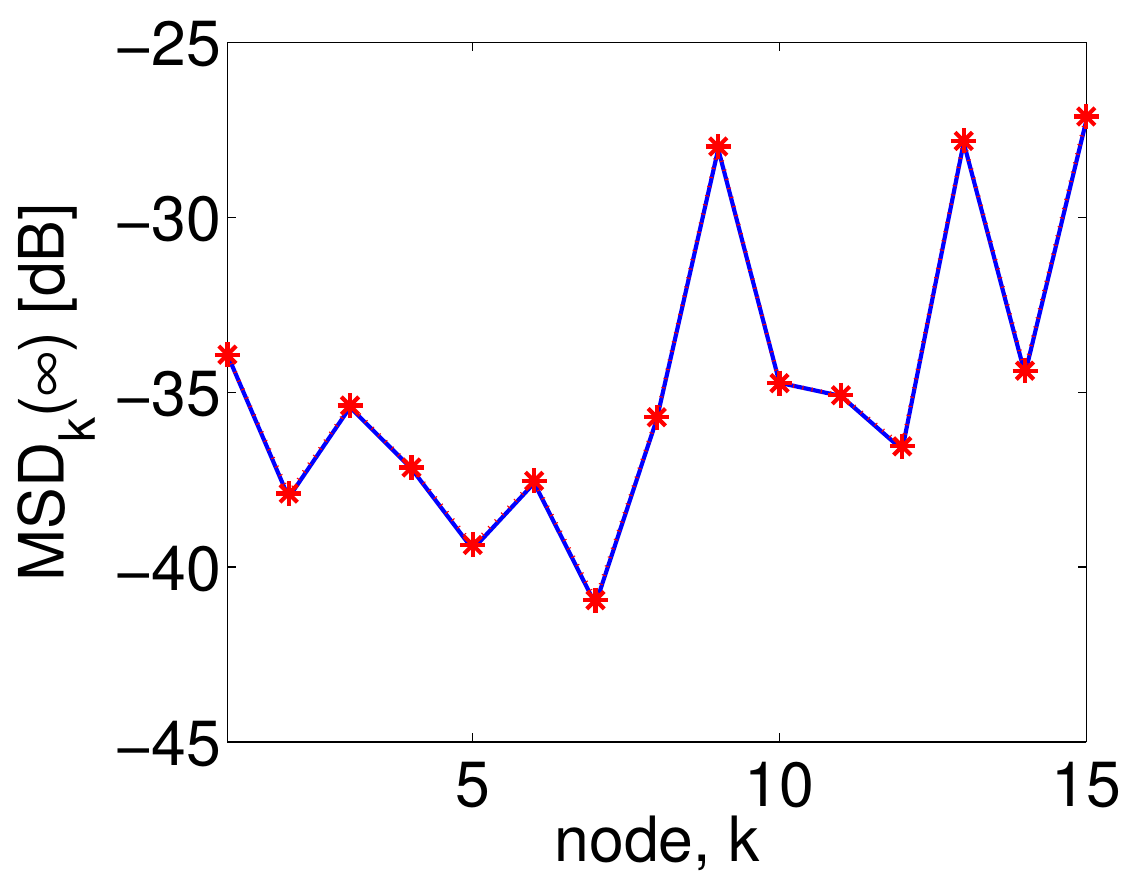} \\
			\footnotesize{Scenario (a)}&\footnotesize{Scenario (b)}\\
			\includegraphics[width=0.4\linewidth]{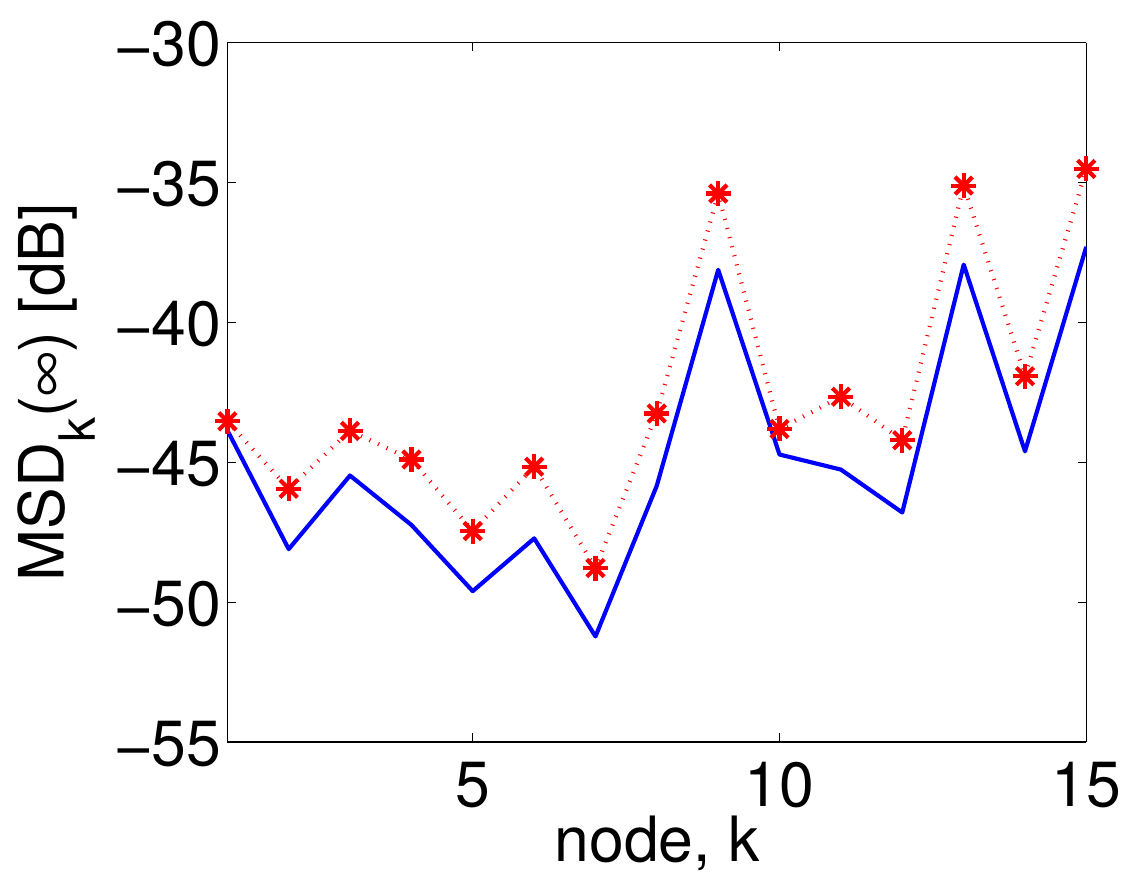} & \includegraphics[width=0.4\linewidth]{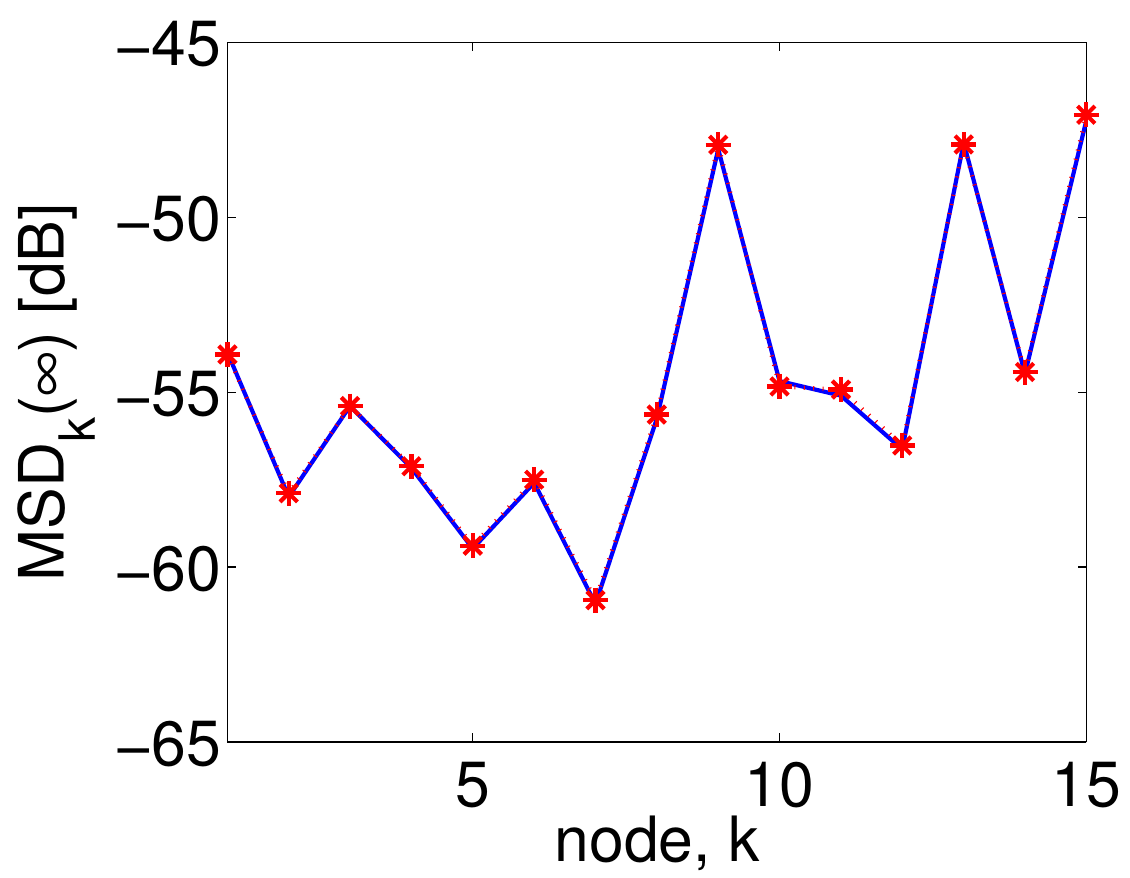} \\
			\footnotesize{Scenario (c)}&\footnotesize{Scenario (d)}\\
        \vspace*{-0.3cm}
		\end{tabular}
		\caption{Comparison between empirical performance and the theoretical model for the node steady-state MSD.}
		\label{fig:model_stationary}
	\vspace*{-0.3cm}
	\end{center}

\end{figure}

In the first place, we carry out some numerical simulations to validate the analysis of Section \ref{sec:performanceanalysis}. To do so, we compare in Fig. \ref{fig:model_stationary} the theoretical and empirical steady-state $\text{MSD}_k$ for the nodes of a D-ATC scheme with Metropolis combiners \cite{sayed2012diffusion} in the stationary estimation scenario. Our objective in this subsection is just to show that the analysis correctly predicts the steady-state performance of each individual node, as well as the $\text{NMSD}$. Although we consider just the case of Metropolis combination rule, we have checked that other rules, e.g., uniform combiners, would lead to similar conclusions about the accuracy of the analysis. In Fig. \ref{fig:model_stationary}, we plot the steady-state MSD for four different scenarios where the step sizes $\mu_{k}$ and the noise variances $\sigma_{v,k}^2$ have been varied from those in Fig. \ref{fig:network}, according to Table \ref{tab:scenarios}.
\begin{table}
	\centering
	\caption{\label{tab:scenarios} Scenarios simulated in Fig. \ref{fig:model_stationary}.  \vspace*{-0.2cm}}
	\begin{tabular}{llll}
      		\toprule Scenario (a) & Scenario (b) & Scenario (c) & Scenario (d) \\
		\midrule $\mu_k$ & $\mu_k/10$ & $\mu_k$ & $\mu_k/10$ \\
		$\sigma_{v,k}^2$ & $\sigma_{v,k}^2$ & $\sigma_{v,k}^2/10$ & $\sigma_{v,k}^2/10$ \\
		\bottomrule
	\end{tabular}
\vspace*{-0.5cm}
\end{table}
From Fig. \ref{fig:model_stationary} we can conclude that the matching between the analysis and the simulation is quite good, even for not so small step sizes [scenarios (a) and (c)]. 

\begin{figure}[t]
	\centering	
	\psfrag{logtrQ}[][c]{\footnotesize{$\log\left(\Tr\{\mathbf{Q}\}\right)$}}
	\includegraphics[width=0.8\linewidth]{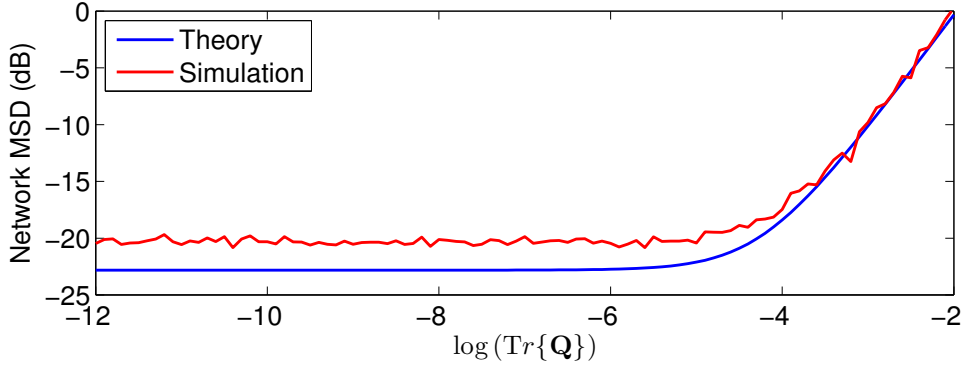}
\vspace*{-0.2cm}
	\caption{Comparison between the analysis and the simulation in a tracking scenario with respect to the logarithm of $\Tr\{\mathbf{Q} \}$.}
	\label{fig:model_track}
\vspace*{-0.3cm}
\end{figure}

We have also studied the accuracy of the model in tracking situations. In Fig. \ref{fig:model_track} we plot the steady-state NMSD for different speeds of change, i.e., values of $\Tr\{\mathbf{Q}\}$. We can see that the matching is also quite good, especially for fast changes. For smaller $\Tr\{\mathbf{Q} \}$ we observe a mismatch up to 2 dB, similarly to the stationary scenario depicted in Fig. \ref{fig:model_stationary}(a).

\subsection{Stationary performance of D-ATC with adaptive combiners }
\label{sec:exp_station}

 \begin{figure}[t]
 	\begin{center}
 		\begin{tabular}{cc}
 			\includegraphics[width=0.8\linewidth]{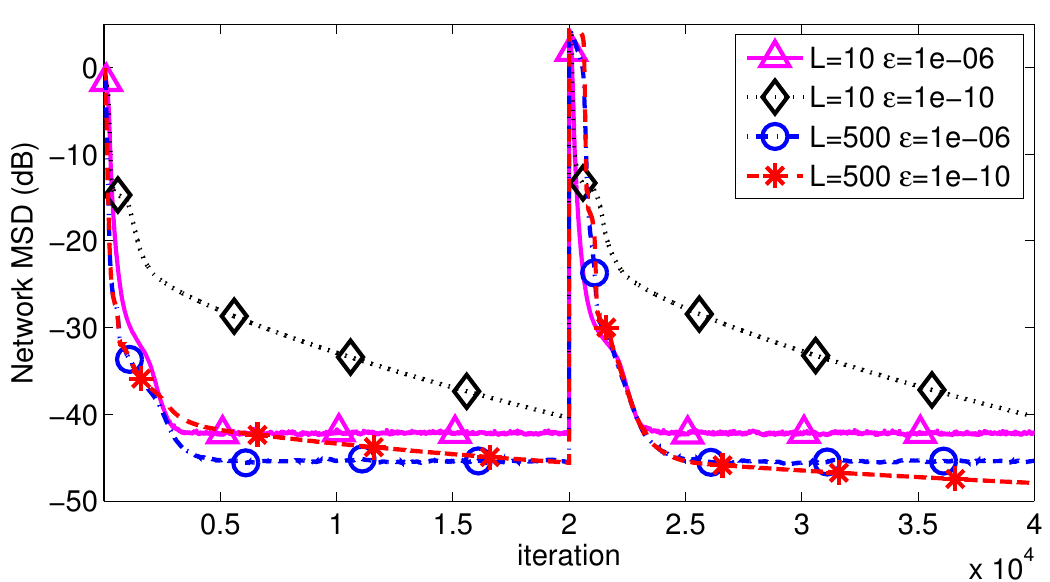}\\
 			(a) \footnotesize{D-ATC with APA adaptive combiners} \\
 			\includegraphics[width=0.8\linewidth]{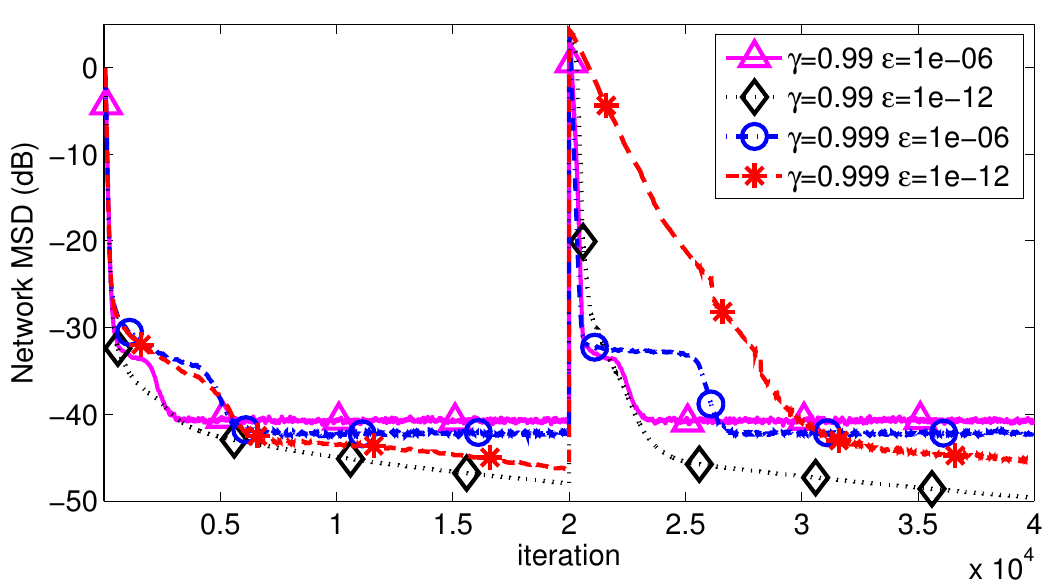}\\
 			(b)\footnotesize{ D-ATC with LS adaptive combiners}\\
         \vspace*{-0.3cm}
 		\end{tabular}
 		\caption{Parameter selection for D-ATC with a) APA and b) LS adaptive combiners.}
 		\label{fig:sim_swept}
 		\vspace*{-0.6cm}
 	\end{center}
 \end{figure}

Before comparing the performance of D-ATC and ATC with adaptive combiners, we study in Fig. \ref{fig:sim_swept} the sensitivity of the proposed combiner learning rules, APA and LS, with respect to their settings. We observe that there is a trade-off between convergence/reconvergence speed and steady-state performance in the selection of these parameters. In fact, we can conclude that the influences of different parameters are coupled among them.

Regarding the forgetting factor $\gamma$ in the LS rule,  note that, when it is correctly chosen [see Fig. \ref{fig:sim_swept}(b)], we can obtain a large steady-state enhancement hardly affecting the convergence. This was not the case with the rectangular window \cite{fernandez2012novel}, where instability issues prevented us from using a very small regularization constant, and limited the number of useful window sizes, causing degradation in the steady-state performance.
%
%
%

\begin{figure}[ht]
	\begin{center}
	\begin{tabular}{c}
		\includegraphics[width=0.8\linewidth]{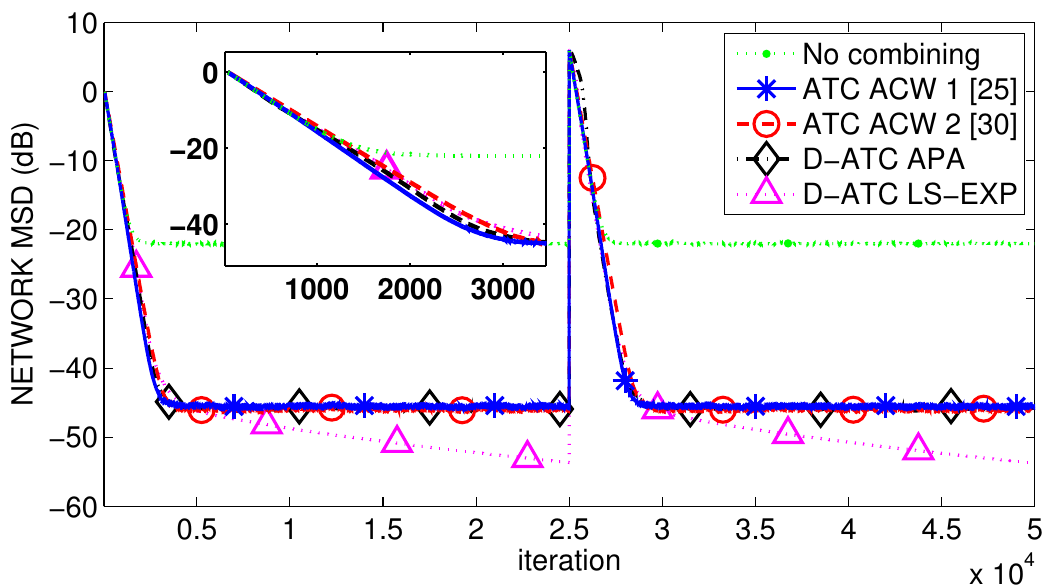}\\
		{\small (a) Homogeneous network}\\
		\includegraphics[width=0.8\linewidth]{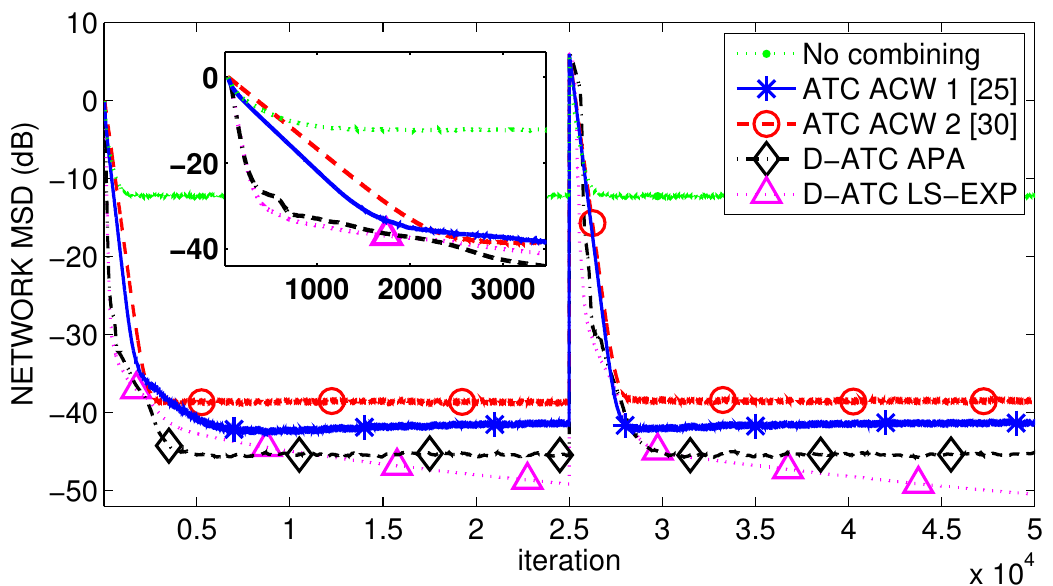}\\
		{\small (b) Heterogeneous network}
	\end{tabular}
	\end{center}
\vspace*{-0.2cm}
\caption{Network MSD performance for a stationary estimation problem. A zoom of the first 3000 iterations is provided for a more clear illustration of the convergence properties of the algorithms.}
\label{fig:sim_station}
\vspace*{-0.3cm}
\end{figure}

Next, we compare our D-ATC scheme with adaptive combiners, with other state-of-the-art ATC algorithms with adaptive combiners: 1) ATC with adaptive combiners proposed by Takahashi \emph{et al.} \cite{Takahashi2010}, and 2) a more recent approach by Tu \emph{et al.} \cite{sayed2012diffusion, tu2011optimal}. We also include a baseline network where the nodes do not combine their estimates. The free parameters of all algorithms are chosen to maximize the steady-state performance while keeping a similar convergence rate and, for reproducibility, are shown in Table \ref{tab:parameters}. {Fig. \ref{fig:sim_station} shows the results for all the mentioned schemes. Although we are more interested in the heterogeneous case, Fig. \ref{fig:sim_station}(a) shows also a comparison for a similar homogeneous network where all the step sizes $\tilde{\mu}_k$ are $0.1$ to show the suitability of adaptive combiners in general.}


\begin{table}[ht]
	\centering
	\caption{\label{tab:parameters} Simulation Parameters.}
\vspace*{-0.2cm}
\begin{tabular}{cllll}
	\toprule  & D-ATC& D-ATC& ATC& ATC\\
	  & APA & LS & ACW 1 \cite{Takahashi2010} & ACW 2 \cite{tu2011optimal} \\
	\midrule \multirow{2}{*}{Stationary.} & $L=500$ & \multirow{2}{*}{$\gamma = 0.99$} & \multirow{2}{*}{$\alpha=0.2$} & \multirow{3}{*}{$\nu=0.1$} \\
	\multirow{2}{*}{Fig \ref{fig:sim_station}}  & $\epsilon=10^{-6}$   &  \multirow{2}{*}{$\epsilon=10^{-12}$}  & \multirow{2}{*}{$\epsilon=10^{-6}$} &  \\
	& $\mu_c=1$   &    &  &  \\
		\hline
		\multirow{2}{*}{Tracking.} & $L=10$ & \multirow{2}{*}{$\gamma = 0.9999$} & \multirow{2}{*}{$\alpha=0.05$} & \multirow{3}{*}{$\nu=0.2$} \\
		\multirow{2}{*}{Fig \ref{fig:sim_track}.(a)} & $\epsilon=10^{-6}$   &  \multirow{2}{*}{$\epsilon=10^{-10}$}  & \multirow{2}{*}{$\epsilon=10^{-6}$} &  \\
		& $\mu_c=1$   &    &  &  \\	
	\hline
	\multirow{2}{*}{Tracking.} & $L=500$ & \multirow{2}{*}{$\gamma = 0.99$} & $\alpha=0.2$& \multirow{3}{*}{$\nu=0.1$} \\
	\multirow{2}{*}{Fig \ref{fig:sim_track}.(b)} & $\epsilon=10^{-6}$   &  \multirow{2}{*}{$\epsilon=10^{-10}$}  & \multirow{2}{*}{$\epsilon=10^{-6}$} &  \\
	& $\mu_c=1$   &    &  &  \\
	\bottomrule
\end{tabular}
\end{table}


{If we compare the results in Figs. \ref{fig:sim_station}(a) and (b), a first conclusion we can extract is that heterogeneous networks achieve a faster initial convergence and after the change in the optimal solution in the middle of the experiment (iteration $2.5 \cdot 10^4$). For the homogeneous networks, it is interesting to notice that D-ATC with LS adaptive combiners can obtain an additional gain in steady state with respect to all other schemes, illustrating the suitability of the network MSE as the optimization criterion to update the combiners.}

{Focusing now on the heterogeneous case, in Fig. \ref{fig:sim_station}(b), we can see that D-ATC with both adaptive rules (APA and LS) significantly outperforms standard ATC both in steady state and during the convergence (see the zoom of the first 3000 iterations for a more clear comparison among algorithms). The combination of our adaptive rules and the decoupled scheme seems to be more effective in this heterogeneous setup. Note that adaptive rules for learning the combination weights for standard ATC \cite{Takahashi2010,tu2011optimal}, are derived for homogeneous networks, when only the noise variance changes among the nodes. That explains most of the gap between both approaches.}


\subsection{Tracking performance of D-ATC with adaptive combiners}

\begin{figure}[t]
\centering
\subfloat[Tracking a fast system, $\Tr\{\mathbf{Q}\}=10^{-4}$.]{\includegraphics[width=0.8\linewidth]{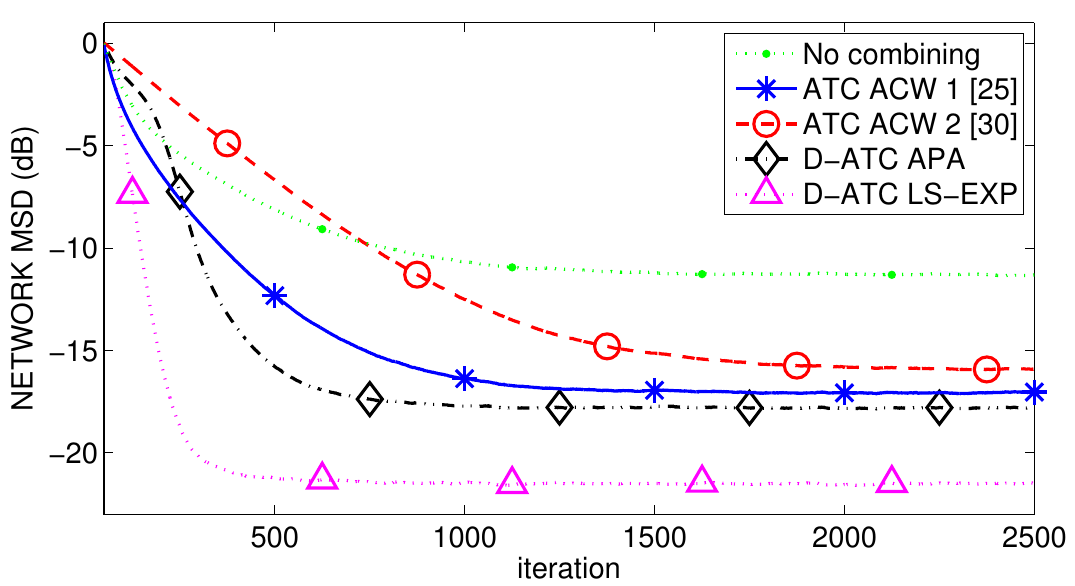}%
\label{fig:track_fast}}
\hfil
\subfloat[Tracking a slow system, $\Tr\{\mathbf{Q}\}=10^{-8}$.]{\includegraphics[width=0.8\linewidth]{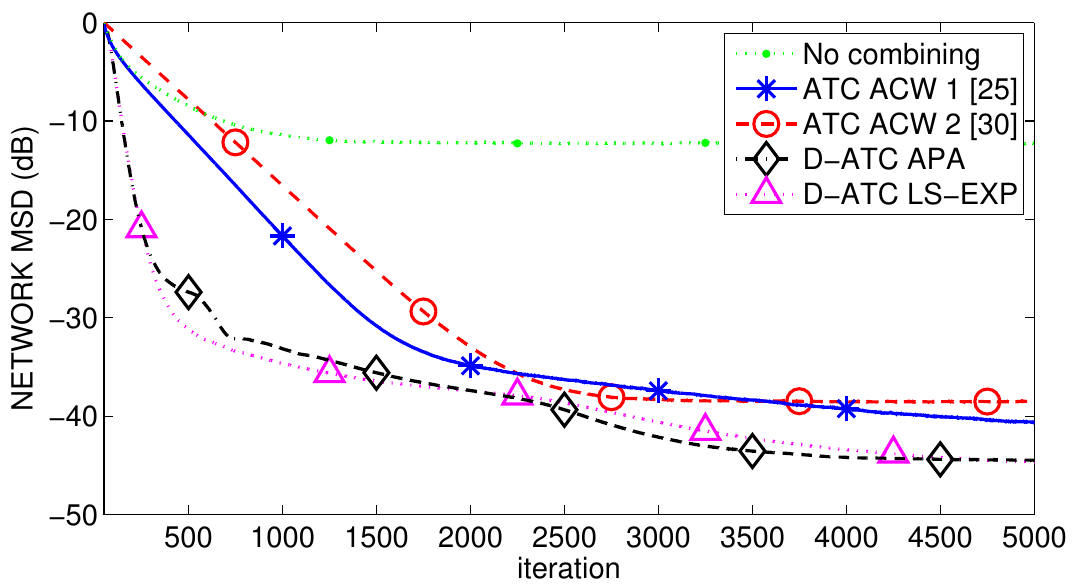}%
\label{fig:track_slow}}
\vspace*{-0.0cm}
\caption{Network MSD performance for a tracking problem: (a) fast variations, $\Tr\{\mathbf{Q}\}=10^{-4}$, (b) slow variations $\Tr\{\mathbf{Q}\}=10^{-8}$.}
\label{fig:sim_track}
\vspace*{-0.3cm}
\end{figure}

We compare in Fig. \ref{fig:sim_track} the performance of D-ATC and ATC, both with adaptive combiners, when tracking a time-varying solution for two different values of $\Tr\{\mathbf{Q}\}$. The parameters of these simulations are shown in Table \ref{tab:parameters}. 
Analyzing the results, we can conclude that D-ATC outperforms both ATC techniques in terms of convergence and steady state, both for the fast and slow time-varying systems.

In conclusion, in all the presented experiments, the proposed D-ATC diffusion scheme outperforms standard ATC, when both schemes use adaptive rules to learn their combiners. 

\subsection{Networks with uninformed nodes}

{In this section, we consider a completely different kind of heterogeneous networks. We follow \cite{Tu_TSP2013} that implements a network with both informed and uninformed nodes (i.e., with and without access to local measurements). In \cite{Tu_TSP2013}, it was shown that ATC networks with fixed combiners do not necessarily benefit from an increased number of informed nodes. In this section we show that our proposed scheme with adaptive combiners is able to use additional data more efficiently, so that an increment in the available information does not degrade network performance.}

\begin{figure}[t!]
\centering
\includegraphics[width=\linewidth]{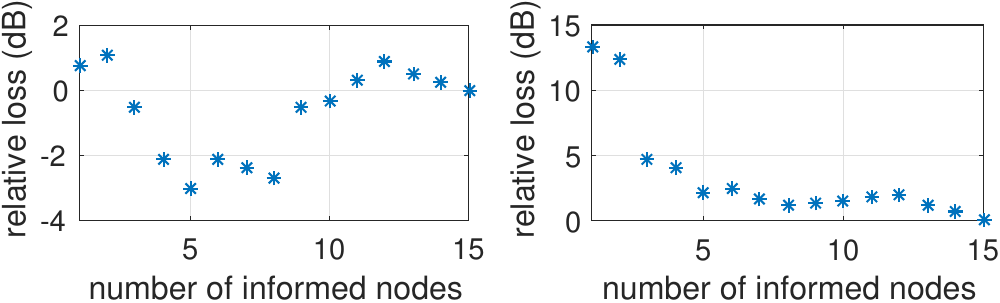}
\begin{tabular}{cc}\footnotesize(a) ATC with uniform combiners \qquad \qquad & \footnotesize (b) D-ATC LS-EXP\end{tabular}
\vspace*{-0.2cm}
\caption{Relative loss in terms of Network MSD performance for a stationary estimation problem in a network with uninformed nodes for ATC with uniform combiners (left) and D-ATC with LS combiners (right). Relative loss is measured with respect to the NMSD of a fully informed network for either case (reference NMSD levels are $-23.5$ dB for the ATC network with static combiners, and $-41.6$ dB for the D-ATC LS scheme).}
\label{fig:sim_uninformed}
\vspace*{-0.3cm}
\end{figure}

{We consider again the network in Fig. \ref{fig:network} when we increase the number of informed nodes (nodes that receive data and perform the adaptation step) from 1 to 15 nodes. In Fig. \ref{fig:sim_uninformed}(a) we show the steady-state NMSD for the standard ATC algorithm with uniform combiners. In such figure, we observe the counterintuitive result of \cite{Tu_TSP2013}: An increment on the number of informed agents in a network can deteriorate its overall performance. However, when varying the number of informed nodes in a D-ATC network with adaptive LS combiners, an increment on the number of informed agents leads to improved performance, since the combiners are able to modify their values to exploit the new available information. This result further justifies the need of using adaptive combiners in heterogeneous networks.}

\subsection{{Computational cost}}

{Finally, we have calculated the computational cost incurred by all the algorithms and configurations evaluated experimentally in this section, as a function of the number of products, sums, divisions and comparisons per iteration.}

{As it can be seen in Table \ref{C_C_EXP}, the performance gains achieved by our proposal have an associated increment in the computational burden with respect to that of the ATC with adaptive combiners in \cite{Takahashi2010} and \cite{tu2011optimal}. This increment results more important for the case of the APA rule with large projection order $L$ than for the case of both the LS algorithm and the APA rule with small $L$, where the computational cost is on the same order of magnitude than for \cite{Takahashi2010} and \cite{tu2011optimal}. {This computational cost could be a limitation in densely connected networks.}

\begin{table}
\centering
\caption{\label{C_C_EXP} Total computational cost for the algorithms and configurations employed in the experiments.}
\vspace*{-0.2cm}
\begin{tabular}{@{}lllll@{}}
\toprule
Operation & Sums & Mult. & Div.& Comp.\\
\midrule
ATC ACW 1 \cite{Takahashi2010} & 22765 &  19878 & 30 & 81   \\
ATC ACW 2 \cite{tu2011optimal} & 13011 &  9873 & 30 & 0   \\
D-ATC LS & 25478 & 25684 & 30 & 81   \\
D-ATC APA ($L=10)$ & 11717 &  11553 & 30 & 81   \\
D-ATC APA ($L=500$) & 76397 & 76233 & 30 & 81   \\
\bottomrule
\end{tabular}
\vspace*{-0.3cm}
\end{table}


\section{Conclusions and Future Work}
\label{sec:conclusion}

{Heterogeneous diffusion networks offer some additional flexibility with respect to homogeneous networks in which all nodes implement the same update rule using common parameters. In this paper, we have presented a novel diffusion scheme that is especially fitted to heterogeneous networks. Each node of our decoupled ATC (D-ATC) scheme keeps a purely local estimate of the solution vector, and calculates an improved combined estimation using its local estimation and combined estimates received from other nodes in the network. We have shown that, if equipped with appropriate schemes for adapting the network combiners, the proposed diffusion scheme can outperform existing ATC networks (both with fixed and adaptive combiners), requiring only a slight increment in the computational cost.}


{This work opens a number of research lines worth exploring. From our point of view, one of the most important is the analysis of asynchronous adaptation in networks. The decoupled nature of D-ATC strategy would make it a good option in such a case. Finally, it is also necessary to evaluate these schemes in the resolution of real tasks. We expect that this contribution helps to further develop the applicability of these networks.}


%

\appendices
\section{Affine Projection Algorithm derivation}
\label{apen:APA}
Consider the cost function defined in \eqref{MSE_apa} and repeated here for convenience
\begin{equation}
\label{MSE_apa_rep}
\text{MSE}_k(n) = \mathbb{E}\left\{ \left[e_k(n)-\bar{\bf{c}}_k^{\T}(n)\tilde{\mathbf{y}}_k(n)\right]^2\right\}.
\end{equation}

Applying the regularized Newton's method \cite{sayed2008adaptive} to minimize \eqref{MSE_apa_rep}, we obtain
\begin{equation}
\label{Newton}
{\bf{\bar{c}}}_{k}(n)={\bf{\bar{c}}}_{k}(n-1)+\mu_c[\epsilon {\bf{I}}_{\bar{N}_k}+{\bf{R}}_{\tilde{\bf{y}}_k}]^{-1}[{\bf{R}}_{e_k,\tilde{\bf{y}}_k}-{\bf{R}}_{\tilde{\bf{y}}_k}{\bf{\bar{c}}}_{k}(n-1)]
\end{equation}
where ${\bf{R}}_{\tilde{\bf{y}}_k}$ is the autocorrelation matrix of vector $\tilde{\bf{y}}_k(n)$, and ${\bf{R}}_{e_k,\tilde{\bf{y}}_k}$ is the cross-correlation vector between $\tilde{\bf{y}}_k(n)$ and $e_k(n)$.

Replacing ${\bf{R}}_{\tilde{\bf{y}}_k}$ and ${\bf{R}}_{e_k,\tilde{\bf{y}}_k}$ by their approximations based on averages over the $L$ most recent values of $\tilde{\bf{y}}_k(n)$ and $e_k(n)$ \cite{sayed2008adaptive}, we obtain the update equation for ${\bf{\bar{c}}}_{k}(n)$ described in \eqref{APA}:
\begin{equation}
\label{APA_rep}
{\bf{\bar{c}}}_{k}(n) ={\bf{\bar{c}}}_{k}(n-1) + \mu_c[\epsilon {\bf{I}}_{\bar{N}_k}+{\bf{\tilde{Y}}}_{k}^{\T}(n){\bf{\tilde{Y}}}_{k}(n)]^{-1} {\bf{\tilde{Y}}}^{\T}_{k}(n) \times 
[{\bf{e}}_k(n)-{\bf{\tilde{Y}}}_{k}(n){\bf{\bar{c}}}_{k}(n-1)].
\end{equation}
\section{Least-Squares Algorithm derivation }
\label{apen:LS}
We start from the cost function \eqref{eq:JkLS}, where we rewrite
\begin{equation}
\check{e}_k(n,i)=e_k(i)+\sum_{\ell=1}^{\bar{N}_k}{c}_{\ell k}(n)\left[y_{k}(i)-y_{\ell k}(i)\right].\label{eq:ekni2}
\end{equation}

Taking now the derivatives of \eqref{eq:JkLS} with respect to each combination
weight $c_{m k}(n)$, with $m = 1, 2, \ldots , \bar{N}_k$, we obtain
\begin{equation}
\frac{\partial J_k(n)}{\partial c_{m k}(n)}=2\sum_{i=1}^{n}\beta(n,i)\check{e}_k(n,i)\left[y_{k}(i)-y_{mk}(i)\right].\label{eq:dJk}
\end{equation}
Replacing \eqref{eq:ekni2} in \eqref{eq:dJk}, setting the result to zero, and after some algebraic manipulations, we obtain
\begin{align}&\sum_{i=1}^{n}\sum_{\ell=1}^{\bar{N}_k}\beta(n,i)c_{\ell k}(n)\tilde{y}_{\ell k}(i)\tilde{y}_{m k}(i) 
=\sum_{i=1}^{n}\beta(n,i)e_k(i)\tilde{y}_{m k}(i).\nonumber\label{eq:normal1}
\end{align}
This defines for for each node $k$ a system with $\bar{N}_k$ equations, introducing the usual matrix notation, reads
\begin{equation}
\mathbf{P}_k(n)\bar{\mathbf{c}}_k(n)=\mathbf{z}_k(n),\label{eq:Pcz}
\end{equation}
where $\mathbf{P}_k(n)$ is a square symmetric matrix of size $\bar{N}_k$ with components
\begin{equation}
\left[\mathbf{P}_k(n)\right]_{p, q}=\sum_{i=1}^{n}\beta(n,i)\tilde{y}_{(\bar{b}_{k}^{(p)},k)}(i)\tilde{y}_{(\bar{b}_{k}^{(q)}, k)}(i),
\end{equation}
with $p,q=1,2,\ldots,\bar{N}_k$. We introduce the index $\bar{b}_{k}^{(p)}$ which is the index of the $p$-th neighbor of $k$. In addition, $\mathbf{z}_k(n)$ is a column vector of length $\bar{N}_k$, whose $p^{\rm th}$ element is given by
\begin{equation}
{z}_k^{(p)}(n)=\displaystyle\sum_{i=1}^{n} \beta(n,i) e_{k}(i)\tilde{y}_{(\bar{b}_{k}^{(p)},k)}(i),
\end{equation}
for $p=1,2,\ldots,\bar{N}_k$.
Thus, the solution of the
problem is obtained from \eqref{eq:Pcz} using Tikhonov method \cite{wahba1977practical}, which leads to \eqref{eq:Pcz2}. 

\vfill
\end{document}